\documentclass[12pt]{iopart}

\pdfoutput=1

\expandafter\let\csname equation*\endcsname\relax
\expandafter\let\csname endequation*\endcsname\relax

\usepackage{graphicx,bm,hyperref,amsmath}

\def\graphicscale{\twocolumn@sw{0.3}{0.4}}
\def\graphicthreescale{\twocolumn@sw{0.3}{0.4}}

\begin{document}

\title{Scaling properties of work fluctuations after quenches at
  quantum transitions}

\author{Davide Nigro, Davide Rossini and Ettore Vicari}
\address{Dipartimento di Fisica dell'Universit\`a di Pisa
        and INFN, Largo Pontecorvo 3, I-56127 Pisa, Italy}

\ead{davide.rossini@unipi.it}
\vspace{10pt}

\begin{indented}
\item[]Authors are listed in alphabetic order
\end{indented}


\begin{abstract}
  We study the scaling properties of the statistics of the work done
  on a generic many-body system at a quantum phase transition of any
  order and type, arising from quenches of a driving control
  parameter.  For this purpose we exploit a dynamic finite-size
  scaling framework.  Namely, we put forward the existence of a
  nontrivial finite-size scaling limit for the work distribution,
  defined as the large-size limit when appropriate scaling variables
  are kept fixed.  The corresponding scaling behaviors are thoroughly
  verified by means of analytical and numerical calculations in two
  paradigmatic many-body systems as the quantum Ising model and the
  Bose-Hubbard model.
\end{abstract}

\vspace{1pc}



\section{Introduction}

Understanding the dynamics of quantum many-body systems is a highly
non trivial task, which is nowadays gaining a great deal of interest,
thanks to the amazing experimental progress in the field of quantum
simulation.  A paradigmatic setup is represented by ultracold atoms
trapped in optical lattices: the system can be manipulated and
controlled, both in space and in time, with an unprecedented accuracy
as compared to any solid-state counterpart, preserving coherence over
long time scales~\cite{BDZ-08}.  In the light of these achievements,
it has become particularly relevant to develop specific theoretical
frameworks that could work in out-of-equilibrium situations and for
strong interactions, overcoming the limitations of linear-response and
perturbation theory.  The ultimate purpose is to shed light on some
fundamental issues of quantum mechanics that have been recently
resurfaced, such as equilibration and thermalization of closed
systems~\cite{PSSV-11, HN-15}, or the emergence of universality in the
dynamics across a critical point~\cite{Dziarmaga-10}.

Important progress has been also made in the intimately related field
of stochastic thermodynamics, where non-equilibrium fluctuation
relations have been put forward, both for
classical~\cite{Jarzynski-11, Seifert-12} and quantum~\cite{CHT-11,
  EHM-09} systems.  The beauty of fluctuation theorems stands on their
completely general validity in out-of-equilibrium conditions, being
able to characterize the full non-linear response of the system to any
(perturbative or not) driving.  These rely on the analysis of
thermodynamic key concepts as work, heat, and entropy, which represent
stochastic variables with definite probability distributions.  We
shall stress that, in the quantum realm, the notion of quantities such
as the work performed on a system are not observables in the usual
sense, but result from an out-of-equilibrium process~\cite{TLH-07,
  TH-16}.

Here we focus on the statistics of the work done on a many-body
system, close to a quantum phase transition, when this is driven out
of equilibrium by suddenly switching one of the control
parameters~\cite{CHT-11, GPGS-18}.  Several issues related to this
topic have been already discussed in a variety of physical
implementations, including spin chains~\cite{Silva-08, DPK-08,
  Dorner-etal-12, Mascarenhas-etal-14, MS-14, ZT-15, SD-15,
  Bayat-etal-16}, fermionic and bosonic systems~\cite{DL-08,
  GS-12,SRH-14, SGLP-14}, quantum field theories~\cite{SGS-13, SS-13,
  PS-14, Palmai-15}, as well as different contexts like cyclically
driven systems~\cite{BAKP-11} and dynamic quantum phase
transitions~\cite{HPK-13}.  It has been also shown that the work
statistics can be experimentally measured in present-day
ultracold-atom systems, by means of ion traps~\cite{HSDL-08} or Ramsey
interferometry~\cite{Dorner-etal-13, MDP-13}.

While the research done so far in this context has mostly addressed
the thermodynamic limit of systems close to criticality (see, e.g.,
Ref.~\cite{GPGS-18} and references therein), in order to achieve a
deep understanding of any reliable quantum-simulation experiment, it
is of primary importance to exploit the impact of having a finite
size.  Studying the finite-size dependence of the work statistics in
nonequilibrium phenomena is also particularly attractive from a
conceptual point of view.  Indeed, for a global quench the work is
extensive.  Therefore in the thermodynamic limit, i.e., large-volume
keeping the Hamiltonian parameter fixed, one expects the probability
associated with the work density $W/V$ (where $V$ is the volume of the
system) to be sharply peaked around its average value, with typical
fluctuations suppressed as $V^{-1/2}$. This suggests that fluctuations
around the work average, and in particular deviations from Gaussian
behaviors, may be only observable for relatively small systems.

The natural theoretical context where to set up the analysis is the
finite-size scaling (FSS) framework, that has been proven to be
effective in proximity of any quantum transition.  Indeed the
emergence of FSS limits has been predicted both for
continuous~\cite{CPV-14} and first-order~\cite{CNPV-14} quantum
transitions (CQTs and FOQTs, respectively), as well as in a dynamic
FSS context (DFSS), to describe the quantum dynamics of finite-size
many-body systems subject to time-dependent
perturbations~\cite{PRV-18b,PRV-18}.  Analogous FSS frameworks have
been recently exploited to study quantum-information based concepts,
as entanglement among parts of the system~\cite{AFOV-08,CCD-09,CPV-14}
as well as other indicators of genuine quantum
correlations~\cite{TRHA-11, DLLS-12, CMC-13, GH-13, Bayat_2017,
  DS-18}, the fidelity~\cite{Gu-10,RV-18}, and
decoherence~\cite{Zurek-03,JP-09,V-18}.  The purpose of this paper is
to extend the analysis to the statistics of the work, and prove its
validity in different paradigmatic quantum many-body systems, which
exhibit both CQTs and FOQTs.

This paper is organized as follows.  In Sec.~\ref{qhprwd} we fix our
setting, by defining the relevant quantities and the quench protocols
that will be considered in the following.  Section~\ref{dfsswd}
contains our general DFSS framework for the work distribution, and
constitutes the core of this work. We specifically address both
continuous (Sec.~\ref{fsscqt}) and first-order (Sec.~\ref{dfsscqt})
transitions, clarifying how the relevant scaling variables can be
defined in the two cases.  Our theory is then verified in the quantum
Ising model (Sec.~\ref{checks}), both at its CQT and along the FOQT
line, and in the hard-core Bose-Hubbard model (Sec.~\ref{xxzmod})
across the vacuum-to-superfluid transition, also generalizing to the
presence of an external trapping potential (Sec.~\ref{tsswf}).
Finally, Sec.~\ref{conclu} is devoted to the conclusions and
perspectives of this research.

\section{Quench protocols and work fluctuations distribution}
\label{qhprwd}

A sudden quench is a protocol which can be generally performed within
a family of Hamiltonians, that are written as the sum of two
noncommuting terms:
\begin{equation}
  H(\lambda) = H_c + \lambda H_p .
  \label{hlam}
\end{equation}
The tunable parameter $\lambda$ enables one to modify the strength of
the {\em perturbation} $H_p$, e.g.,~a magnetic field term in a system
of interacting spins, with respect to the {\em unperturbed}
Hamiltonian $H_c$.  We denote with $|n_\lambda \rangle$ the
eigenstates of energy $E_{n}^{\lambda}$ of $H(\lambda)$ (we assume a
discrete spectrum, as it is generally appropriate for finite-size
systems), in particular $|0_\lambda \rangle$ is the corresponding
ground state.  The idea of a quantum quench is to prepare the system
in the ground state of the Hamiltonian~(\ref{hlam}) for a given value
$\lambda_0$ of the $\lambda$-parameter, or in the corresponding
equilibrium thermal state defined by the Gibbs distribution
\begin{equation}
  p_n^{\lambda_0}= \frac{e^{-\beta
      E_{n}^{\lambda_0}}}{Z(\lambda_0)},\qquad Z(\lambda_0) = \sum_m
  e^{-\beta E_m^{\lambda_0}}.
  \label{gibbsdistr}
\end{equation}
Here $\beta \equiv 1/T$ denotes the inverse temperature of a given
heat reservoir, with which the system is initially coupled and in
equilibrium (hereafter we will adopt units of $\hbar = k_B = 1$).  At
time $t=0$, the system-bath coupling is switched off and the
$\lambda$-parameter is suddenly changed to $\lambda\neq\lambda_0$.
One is then interested in studying the properties of the system, which
unitarily evolves according to the post-quench Hamiltonian
$H(\lambda)$, after a time $t$. In the case the initial condition is
just the ground state $|0_{\lambda_0}\rangle$ of $H(\lambda_0)$, or
equivalently the zero-temperature limit of the Gibbs distribution, the
resulting dynamic problem corresponds to that of the pure-state
quantum evolution
\begin{equation}
  |\Psi(t)\rangle = e^{-i H(\lambda) t}|0_{\lambda_0}\rangle,
  \label{psit}
\end{equation}
with $|\Psi(t=0)\rangle =|0_{\lambda_0}\rangle$~\cite{PSSV-11}.

The quantum work $W$ associated with such out-of-equilibrium dynamic
protocol, i.e., the work done on the system by quenching the control
parameter $\lambda$, does not generally have a definite value.  More
specifically, this quantity can be defined as the difference of two
projective energy measurements~\cite{CHT-11}.  The first one at $t=0$
projects onto the eigenstates of the initial Hamiltonian
$H(\lambda_0)$ with a probability given by the equilibrium Gibbs
distribution. Then the system evolves, driven by the unitary operator
$U(t,0)=e^{-i H(\lambda) t}$, and the second energy measurement
projects onto the eigenstates of the Hamiltonian $H(\lambda)$. The
work probability distribution can thus be written
as~\cite{CHT-11,TH-16,TLH-07}:
\begin{equation}
  P(W) \equiv P(W, T, \lambda_0, \lambda)
  = \sum_{n,m} \delta \big[ W-(E_n^\lambda-E_m^{\lambda_0}) \big] \,
  \big| \langle n_{\lambda} | m_{\lambda_0} \rangle \big|^2 \,
  p_m^{\lambda_0}.\quad
  \label{pwdefft}
\end{equation}
One may also introduce a corresponding characteristic
function~\cite{Silva-08,CHT-11}
\begin{equation}
C(s)  = \int dW \, e^{i s W} \, P(W),
\label{gudef}
\end{equation}
encoding full information of the work statistics. 

The work distribution~(\ref{pwdefft}) satisfies the quantum version
of the Crooks fluctuation relation~\cite{CHT-11}
\begin{equation}
  \frac{P(W, T, \lambda_0, \lambda) }{P(-W, T, \lambda, \lambda_0)}
  = e^{\beta W} e^{-\beta [ F(\lambda) - F(\lambda_0)]},
  \label{eq:Crooks}
\end{equation}
where the probability distribution in the denominator corresponds to
an inverted quench protocol, from $\lambda$ to $\lambda_0$.  It also
satisfies the Jarzynski equality~\cite{Jarzynski-11, CHT-11} [which
  can be directly derived from Eq.~(\ref{eq:Crooks}), as well]:
\begin{equation}
  \langle e^{-\beta W} \rangle \equiv \int dW\, e^{-\beta W}\, P(W) 
  = e^{-\beta [F(\lambda)-F(\lambda_0)]},
  \label{jeq}
\end{equation}
where $F(\lambda) = - T \,\ln Z(\lambda)$ is the free energy
associated with the Hamiltonian parameter $\lambda$.  Let us also
define the so-called dissipative work~\cite{CHT-11}
\begin{equation}
W_i = W - [F(\lambda)-F(\lambda_0)],
\label{disswo}
\end{equation}
satisfying the inequality $\langle W_i \rangle \ge 0$, which can be
straightforwardly derived using Jensen's inequality.

The zero-temperature limit corresponds to a quench protocol from the
ground state of $H(\lambda_0)$.  Assuming that it is nondegenerate,
the work probability~(\ref{pwdefft}) reduces to
\begin{equation}
  P(W) = \sum_{n} \delta \big[ W-(E_n^\lambda-E_0^{\lambda_0}) \big]
  \; \big| \langle n_{\lambda} | 0_{\lambda_0} \rangle \big|^2,
  \label{pwdef}
\end{equation}
and the dissipative work~(\ref{disswo}) simplifies to
\begin{equation}
W_i= W - (E_0^{\lambda}-E_0^{\lambda_0}).
\label{wit0}
\end{equation}
In this case, the characteristic function $C(s)$ can be simply
written as the amplitude
\begin{equation}
  C(s) = \langle 0_{\lambda_0} | e^{-i H(\lambda_0) s}
  \,e^{iH(\lambda) s} | 0_{\lambda_0} \rangle,
  \label{csamp}
\end{equation}
whose absolute value is related the so-called Loschmidt echo $Q(t) = -
\ln | \langle 0_{\lambda_0} | \Psi(t)\rangle|^2$, providing
information on the overlap between the initial state
$|0_{\lambda_0}\rangle$ and the evolved quantum state
$|\Psi(t)\rangle$ at time $t$, cf. Eq.~(\ref{psit}).

It is even possible to devise generalized time-dependent protocols,
such as those starting from the ground state at $\lambda_0$, and then
evolving the system by changing the protocol parameter $\lambda_p(t)$
with a generic function of the time, up to a final value $\lambda$.
In particular we may write it as
\begin{equation}
\lambda_p(t) = \lambda_0 + (\lambda - \lambda_0) \, f_p(t/t_p), 
\label{gpf}
\end{equation}
so that $f_p(0) =0$ and $f_p(1)=1$, and $t_p$ is the time scale of the
variation.  For example, one may consider a linear protocol $f_p(x) =
x$.  In this more general case, the work distribution~(\ref{pwdefft})
reads~\cite{CHT-11}:
\begin{equation}
  P(W;f_p) = \sum_{n,m} \delta \big[
    W-(E_n^{\lambda}-E_m^{\lambda_0}) \big] \,
  \big| \langle n_{\lambda} | U(t_p;f_p)|m_{\lambda_0} \rangle \big|^2
  \; p_m^{\lambda_0} ,
  \label{pwdefg}
\end{equation}
where $U(t_p;f_p)$ is the evolution operator from $t=0$ to $t_p$
associated with the time variation of Eq.~(\ref{gpf}).  Of course, the
general features of the dynamic behavior must somehow depend on the
protocol function $f_p(x)$.  In the limit $t_p\to 0$, we recover the
sudden-quench protocol. Note also that, in the case of finite-size
systems, even at quantum transitions, large time scales $t_p$ must
eventually give rise to quasi-static adiabatic evolutions, with
vanishing dissipative work, i.e., $\langle W \rangle = F({\lambda}) -
F({\lambda_0})$.

Let us finally comment on another out-of-equilibrium protocol based on
a double sudden quench forming a cycle, similar to that discussed in
Ref.~\cite{HPK-13}.  The quantum evolution starts from an equilibrium
Gibbs distribution at a given temperature $T$, associated with the
Hamiltonian parameter $\lambda_0$.  The $\lambda$-parameter is then
suddenly changed to $\lambda$, at $t=0$. After a time $t$, the
parameter is suddenly quenched back to the value $\lambda_0$. The work
distribution, associated with two projective energy measurements at
$t=0$ and after the second quench at time $t$, can be written as
\begin{equation}
  P(W) = \sum_{n,m} \delta \big[ W-(E_n^{\lambda_0}-E_m^{\lambda_0})
    \big] \, \big| \langle n_{\lambda_0} |
  e^{-iH(\lambda)t} | m_{\lambda_0} \rangle \big|^2
  \;p_m^{\lambda_0}. \label{pwded}
\end{equation}
Of course, in such case we have a further dependence on the time
interval $t$ in which the system evolves under the Hamiltonian
$H(\lambda)$.  The corresponding Jarzynski identity reads as $\langle
e^{-\beta W} \rangle = 1$, since the double-quench protocol provides
an out-of-equilibrium cycle returning to the initial value $\lambda_0$
of the $\lambda$-parameter.  The average work satisfies the
thermodynamic inequality $\langle W \rangle \ge 0$, which can be
easily derived using Jensen's inequality.

In the following sections we characterize the scaling behavior of the
work fluctuations arising from the above quench protocols performed
within the {\em critical} regime of a quantum transition. This
requires that $H_c$ describes a system either at a CQT or a FOQT, and
that the temperature $T$ is sufficiently low.  Moreover, the
perturbation $\lambda H_p$, driving the quench protocol, must be
sufficiently small to maintain the system within the critical
regime. In particular, we derive the scaling behaviors arising from
the interplay of the temperature and the Hamiltonian parameters with
the finite size $L$ of the system.  For this purpose, we are going to
exploit an appropriate DFSS framework~\cite{PRV-18b, V-18}.

\section{Dynamic finite-size scaling of the work distribution}
\label{dfsswd}

\subsection{General Ansatz}
\label{genans}

Let us start by considering a simple quench protocol, driven by the
sudden change of the $\lambda$-parameter of the many-body
Hamiltonian~(\ref{hlam}) at a quantum transition.  We shall see that
the DFSS framework allows us to study the interplay among the
temperature $T$, the quench parameters $\lambda_0,\,\lambda$, and the
finite linear size $L$, assuming that the temperature is sufficiently
small, and that both the initial ($\lambda_0$) and final ($\lambda$)
parameters keep the system close to the transition point.  The
transition point is located at $\lambda=0$, where the Hamiltonian is
exactly $H_c$, while $\lambda$ is a relevant parameter driving the
transition.

Our working hypothesis is based on the existence of a nontrivial DFSS
limit for the work distribution $P(W)$, defined as the large-size
limit keeping the appropriate scaling variables fixed.  Namely, we
conjecture that, at both CQTs and FOQTs, the DFSS of the work
distribution can be written as
\begin{equation}
P(W,T,\lambda_0,\lambda,L) \approx \Delta(L)^{-1} \:
{\cal  P}(\omega,\tau,\kappa_0,\kappa).
\label{genpwsca}
\end{equation}
Here 
\begin{equation}
\Delta(L) = \Delta(\lambda=0,L),
\label{deltaldef}
\end{equation}
where $\Delta(\lambda,L)$ is the energy gap of the lowest states,
while ${\cal P}(\omega,\tau,\kappa_0,\kappa)$ is a function of the
scaling variables
\begin{equation}
\omega = \Delta(L)^{-1} \, W,\qquad \tau = \Delta(L)^{-1} \, T,
\label{ometau}
\end{equation}
$\kappa_0$ and $\kappa$. The latter are other appropriate scaling
variables proportional to $\lambda_0$ and $\lambda$, respectively (see
below). The DFSS limit is defined as the large-size $L\to\infty$ limit,
keeping the scaling variables $\omega$, $\tau$, $\kappa_0$ and
$\kappa$ fixed.

The DFSS of the work distribution in Eq.~(\ref{genpwsca}) allows us to
infer the scaling behavior of the average of the work and its higher
moments, as well:
\begin{equation}
\langle W^k \rangle  
\equiv \int dW\, W^k\, P(W) \approx 
\Delta(L)^k \,{\cal W}_k(\tau,\kappa_0,\kappa), \label{scalwk}
\end{equation}
with
\begin{equation}
{\cal W}_k(\tau,\kappa_0,\kappa)=
\int d\omega\,\omega^k\,{\cal P}(\omega,\tau,\kappa_0,\kappa) .
\end{equation}
Of course ${\cal W}_{k=0}=1$, corresponding to the normalization
condition
\begin{equation}
\int dW P(W) = \int d\omega\, {\cal P}(\omega) = 1.
\label{normpw}
\end{equation}

In the case of the work distribution~(\ref{pwded}) associated with a
double quench protocol in which the $\lambda$-parameter is brought
back to its initial value $\lambda_0$ after a time $t$, we must add a
further scaling variable
\begin{equation}
  \theta_t = \Delta(L) \, t,
  \label{thetadef}
\end{equation}
to take into account the dependence on the time interval of the
evolution of the system under the Hamiltonian $H(\lambda)$.  Scaling
arguments analogous to those applied to the standard quench
protocols~\cite{PRV-18b} lead to the general DFSS ansatz
\begin{equation}
  P(W,T,\lambda_0,\lambda,t,L) \approx \Delta(L)^{-1}\, {\cal
    P}(\omega,\tau,\kappa_0,\kappa,\theta_t),
  \label{genpwscad}
\end{equation}
at both CQTs and FOQTs.

As we shall see later, different scaling behaviors of the work
distribution are expected to occur at CQTs and FOQTs, due to the
different size dependence of the gap $\Delta(L)$ and the FSS variables
$\kappa_0,\,\kappa$: CQTs are generally characterized by power laws
related to the universal critical exponent of the corresponding
universality class, while exponential behaviors emerge at FOQTs.
In the following we discuss more in detail the DFSS frameworks
associated with CQTs and FOQTs, emphasizing their own peculiarities.

\subsection{Continuous quantum transitions}
\label{fsscqt}

The FSS theory at CQTs is well established (see,
e.g.,~Refs.~\cite{CPV-14, Barber-83, Privman-90} and references
therein).  The energy differences $\Delta_i(L)=E_i-E_0$ of the lowest
states [and, in particular, of the ground-state gap $\Delta(L)\equiv
  \Delta_1(L)$] at the transition point behave as
\begin{equation}
  \Delta_i(L) \sim L^{-z},
\end{equation}
where $z$ is a universal dynamic exponent.
The FSS variable associated with the relevant scaling
variable is
\begin{equation}
  \kappa = L^{y_\lambda} \lambda,
  \label{eq:KappaCQT}
\end{equation}
where $y_\lambda$ is a universal critical exponent, given by the
renormalization-group (RG) dimension of the $\lambda$ parameter.  The
equilibrium FSS limit of a generic observable $O$, obtained by taking
$L\to\infty$ keeping $\kappa$ and $\tau \sim L^z \, T$ fixed [see
  Eq.~(\ref{ometau})], is expected to be
\begin{equation}
O(T,\lambda,L) \approx  L^{-y_o} \, {\cal O}(\tau,\kappa),
  \label{cqteq}
\end{equation}
where $y_o$ is the RG dimension of the observable $O$, and ${\cal O}$
is a universal FSS function depending on the geometry of the system
and the type of boundary conditions.

Out-of-equilibrium time-dependent processes also require an
appropriate rescaling of the time $t$, encoded by the scaling variable
$\theta_t \sim L^{-z} t$ [see Eq.~(\ref{thetadef})].  The
corresponding DFSS limit is thus defined, in an analogous way, as the
infinite-volume $L\to\infty$ limit keeping the scaling variables
$\theta_t$, $\tau$, $\kappa$, and $\kappa_0 =L^{y_\lambda} \lambda_0$
fixed.  Then a generic observable $O$ in the DFSS limit is expected to
behave as~\cite{PRV-18b}
\begin{equation}
  O(T,\lambda_0,\lambda,t,L) \approx 
  L^{-y_o} \, {\cal O}(\tau,\kappa_0,\kappa,\theta_t),
  \label{cqtoq}
\end{equation}
where ${\cal O}$ is a DFSS function.

The DFSS asymptotic behavior of the work probability distribution,
given by Eq.~(\ref{genpwsca}), can be thus rewritten after defining a
further scaling variable $\omega \sim L^z \,W$ [see again
  Eq.~(\ref{ometau})], such that
\begin{equation}
  P(W,T,\lambda_0,\lambda,L) \approx L^z \,{\cal
    P}(\omega,\tau,\kappa_0,\kappa).
  \label{Psca}
\end{equation}
Correspondingly, its characteristic function $C(s)$ is expected to
behave as
\begin{equation}
C(s,T,\lambda_0,\lambda,L) \approx {\cal C}(\theta_s,\tau,
\kappa_0,\kappa),
\label{gtsca}
\end{equation}
where $\theta_s = \Delta(L) \, s \sim L^{-z}\,s$, and
\begin{equation}
{\cal C}(\theta_s,\tau,\kappa_0,\kappa) = \int d\omega \,
e^{i\theta_s\omega} \,{\cal P}(\omega,\tau,\kappa_0,\kappa).
\label{gupw}
\end{equation}
Note that this DFSS is consistent with that expected for typical
Loschmidt echos, along the quantum evolution after a quench
protocol~\cite{PRV-18b}. Indeed, due to Eq.~(\ref{csamp}), the
variable $s$ can be naturally considered as a time-like variable,
explaining the time-like scaling of the corresponding FSS variable
$\theta_s$ [see Eq.~(\ref{thetadef})].
One may also derive nontrivial relations analogous to those of
Eqs.~(\ref{eq:Crooks}) and~(\ref{jeq}) in terms of scaling functions
and scaling variables only, using the scaling equation of the
equilibrium free energy~\cite{CPV-14}.

The approach to the asymptotic DFSS behavior is expected to be
characterized by power-law suppressed corrections, typical of general
CQTs~\cite{CPV-14}. In particular, we expect the existence of
$O(L^{-\omega})$ corrections, where $\omega$ is the universal exponent
associated with the leading irrelevant perturbation at the
corresponding fixed point. Such corrections generally appear in the
equilibrium critical behavior of correlations of local
observables. The presence of boundaries generally gives rise to
$O(L^{-1})$ corrections, while in the absence of boundaries, like
periodic boundary conditions, these corrections are absent.  Moreover,
for complex nonlocal quantities, such as the entanglement entropy
between spatial regions, other peculiar power-law corrections may
arise (see, e.g., Ref.~\cite{PRV-18b} and references therein).

Using the general DFSS of the work probability, it is easy to derive
the DFSS of the average work $\langle W \rangle \approx \Delta(L) \,
{\cal W}_1(\tau,\kappa_0,\kappa)$.  This becomes, in the
zero-temperature limit,
\begin{equation}
  \langle W \rangle \approx \Delta(L) \,{\cal W}_1(\kappa_0,\kappa).
  \label{wscacot0}
\end{equation}

We shall see here that the DFSS in Eq.~(\ref{wscacot0}) can be
supported by an alternative derivation.  The average work injected
into the system by a quench from $\lambda_0$ to $\lambda$ is given by
the expectation value of the post-quench Hamiltonian on the initial
(pre-quench) state:
\begin{equation}
  \langle W \rangle = \langle 0_{\lambda_0} | H(\lambda) -
  H(\lambda_0)| 0_{\lambda_0} \rangle = (\lambda - \lambda_0) \langle
  0_{\lambda_0} | H_p | 0_{\lambda_0} \rangle.
  \label{energy}
\end{equation}
It is easy to check that this equation corresponds to the integral
$\langle W \rangle = \int dW\,W\,P(W)$, inserting $P(W)$ as given in
Eq.~(\ref{pwdef}).

In the DFSS limit, we can exploit the equilibrium FSS behavior,
cf. Eq.~(\ref{cqteq}), to evaluate the matrix element $\langle
0_{\lambda_0} | H_p | 0_{\lambda_0} \rangle$.  Assuming that $H_p =
\sum_{\bm x} P_{\bm x}$ is a sum of local terms, we have
\begin{equation}
  \langle W \rangle \approx  L^{d-y_p} \,
  (\lambda-\lambda_0)\, f_p(\kappa_0) ,
  \label{calla2}
\end{equation}
where $y_p$ and $f_p$ are, respectively, the RG dimension and the
equilibrium FSS function associated with the observable $H_p/L^{d}$
(here $d$ is the dimensionality of the system).  Taking into account
the relation~\cite{Sachdev-book}
\begin{equation}
  y_p + y_\lambda = d+z
  \label{ypyl}
\end{equation}
between the RG dimensions of $\lambda$ and of the associated
perturbation $H_p$, Eq.~(\ref{calla2}) can be eventually written as
\begin{equation}
  \langle W\rangle \approx  L^{-z} (\kappa-\kappa_0) \, f_p(\kappa_0),
  \label{calla3}
\end{equation}
in agreement with Eq.~(\ref{wscacot0}), identifying 
\begin{equation}
  {\cal W}_1(\kappa_0,\kappa) \sim (\kappa-\kappa_0)f_p(\kappa_0).  
  \label{w1co}
\end{equation}
This provides a relation between the DFSS function of the average work
and the equilibrium FSS function of the expectation value of the
Hamiltonian term $H_p$ associated with the driving parameter
$\lambda$.

We may also consider the large-volume limit of the above scaling
behaviors.  For $L\to\infty$, the average work is expected to grow as
the volume, which implies
\begin{equation}
  f_p(\kappa_0) \sim |\kappa_0|^{y_p/y_\lambda}, \qquad |\kappa_0|
  \to \infty, \qquad \langle W\rangle \sim L^{d}\,
  (\lambda-\lambda_0) \,|\lambda_0|^{y_p/y_\lambda}.
\label{calla2li}
\end{equation}
Note also that we may write it as
\begin{equation}
  {\langle W\rangle\over L^{d} } \sim \xi_0^{-(d+z)} \delta_\lambda,
  \label{altivsca}
\end{equation}
where $\xi_0 \sim |\lambda_0|^{-1/y_\lambda}$ represents an
infinite-volume correlation length associated with the initial ground
state of $H(\lambda_0)$, and $\delta_\lambda = \lambda/\lambda_0 - 1 =
\kappa/\kappa_0 - 1$.

We stress that the DFSS relations derived above are quite general, in
that they can be applied to any CQT, using the appropriate critical
exponents associated with the corresponding universality class.
Moreover, analogous considerations apply to local quenches, which can
be described by a DFSS as well~\cite{PRV-18}.

We finally mention that the work distribution in the infinite volume
limit is expected to approach a quasi-Gaussian distribution around the
average value of the work density $W/L^d$ with $O(L^{-d/2})$
fluctuations, which is expected to have the general form $P(W) \sim
\exp[-L^d I(W/L^d)]$ with $I(x)\ge 0$~\cite{GS-12}.

\subsection{First-order quantum transitions}
\label{dfsscqt}

Let us now extend the above analysis to FOQTs.  As shown by earlier
works~\cite{CNPV-14, CPV-15, PRV-18c}, isolated many-body
systems at FOQTs develop FSS behaviors as well.  However, they
significantly depend on the type of boundary conditions, in particular
whether they favor one of the phases or they are neutral, giving rise
to FSS characterized by exponential or power-law behaviors.

FOQTs generally arise from level crossings. However level crossings
can only occur in the infinite-volume limit (in the absence of
particular conservation laws).  In a finite system, the presence of a
nonvanishing matrix element among these states lifts the degeneracy,
giving rise to the phenomenon of avoided level crossing.  Here the FSS
is controlled by the energy difference of the avoiding levels, in
particular by the gap $\Delta(L)$ of Eq.~(\ref{deltaldef}). The
appropriate FSS variables are generally given by those in
Eq.~(\ref{ometau}), and by~\cite{CNPV-14}
\begin{equation}
  \kappa = \Delta(L)^{-1} \, E_\lambda(\lambda,L) ,
  \label{kappafoqt}
\end{equation}
$E_\lambda$ being the energy variation associated with the $\lambda$
term (we assume $E_\lambda= 0$ at the transition point).  The DFSS
limit is again defined by the large-$L$ limit, keeping $\omega$,
$\tau$, and $\kappa$ fixed.  Note that the FOQT scenario based on the
avoided crossing of two levels is not realized for any boundary
condition~\cite{CNPV-14}: in some cases the energy difference
$\Delta(L)$ of the lowest levels may even show a power-law dependence
on $L$.  However, the scaling variables $\kappa$ obtained using the
corresponding $\Delta(L)$ turn out to be appropriate as
well~\cite{CNPV-14}.

Similarly to CQTs, the emergence of a DFSS after an out-of-equilibrium
quench protocol $\lambda_0 \to \lambda$ is also expected at FOQTs.
The scaling arguments of Ref.~\cite{PRV-18b} allow us to identify the
additional variables $\kappa_0 = \Delta(L)^{-1}
E_\lambda(\lambda_0,L)$ and $\theta_t=\Delta(L) t$ [see
  Eq.~(\ref{thetadef})].
Using arguments analogous to those at CQTs, we arrive at the DFSS of
the work probability distribution given in Eq.~(\ref{genpwscad}).
These considerations can be straightforwardly extended to generalized
quench protocols, such as those introduced in Sec.~\ref{qhprwd}.

The rest of the paper is dedicated to an explicit verification of the
DFSS that we put forward for the work distribution after a quench in
two paradigmatic examples of quantum many-body systems exhibiting
transitions of different types and order.  Namely, we focus on the
quantum Ising model (Sec.~\ref{checks}) and on the Bose-Hubbard model
(Sec.~\ref{xxzmod}).  We employ both numerical diagonalization
techniques and analytical approaches, when possible.

\section{Results for the quantum Ising model}
\label{checks}

\subsection{The model}
\label{qism}

The Hamiltonian of the $d$-dimensional quantum Ising model defined on
a lattice with $L^d$ sites, in the presence of both a transverse and a
longitudinal field, is given by:
\begin{equation}
  H_{\rm Is} = - J \, 
  \sum_{\langle {\bf x},{\bf y}\rangle} \sigma^{(3)}_{\bf x}
  \sigma^{(3)}_{\bf y} - g\, \sum_{\bf x} \sigma^{(1)}_{\bf x}
  - \lambda \sum_{\bf x} \sigma^{(3)}_{\bf x}.
  \label{hisdef}
\end{equation}
Here $\sigma^{(k)}$ denotes the spin-$1/2$ Pauli matrices ($k=x,y,z$),
the first sum is over all bonds connecting nearest-neighbor sites
$\langle {\bf x},{\bf y}\rangle$, while the other sums are over the
sites.  We assume $J=1$ (ferromagnetic couplings) and $g>0$.  
The phase diagram of these kinds of systems (in various dimensions) 
are known, and present both CQTs and FOQTs.

At $g=g_c$ and $\lambda=0$ (in one dimension, $g_c=1$), the model
undergoes a CQT belonging to the $(d+1)$-dimensional Ising
universality class~\cite{Sachdev-book, ZJ-book, PV-02}, separating a
disordered phase ($g>g_c$) from an ordered ($g<g_c$) one.  The CQT at
$g=g_c$ is characterized by the presence of two relevant Hamiltonian
parameters.  They are $r\equiv g-g_c$ and $\lambda$ (such that they
vanish at the critical point), with RG dimension $y_r$ and
$y_\lambda$, respectively.
The equilibrium critical exponents $y_r$ and $y_\lambda$ are those of
the $(d+1)$-dimensional Ising universality class. For one-dimensional
systems, they are $y_r=1/\nu=1$ and $y_\lambda = (d+3-\eta)/2=
(4-\eta)/2$ with $\eta=1/4$.  For two-dimensional models, they are not
known exactly, but there are very accurate estimates, see, e.g.,
Refs.~\cite{PV-02, Hasenbusch-10, KPSV-16, KP-17}; in
particular~\cite{KPSV-16} $y_r=1/\nu$ with $\nu=0.629971(4)$ and
$y_\lambda = (5-\eta)/2$ with $\eta=0.036298(2)$.  For
three-dimensional systems, they assume the mean-field values $y_r=2$
and $y_\lambda=3$, apart from logarithms.  The temperature $T$ gives
rise to a further relevant perturbation at CQTs; the corresponding
scaling dimension is provided by the dynamic exponent $z=1$ (for any
spatial dimension) characterizing the behavior of the energy
differences of the lowest-energy states, and, in particular, the gap
$\Delta\sim \xi^{-z}$ where $\xi$ is the diverging length scale at the
transition point. Scaling corrections are generally controlled by the
leading irrelevant perturbation, which gives rise to $O(\xi^{-\omega})$
corrections to the asymptotic behavior, and the corresponding
universal exponent is given by $\omega=2$ for $d=1$~\cite{CHPV-00} and
$\omega=0.830(2)$ for $d=2$~\cite{KPSV-16}.

For any $g<g_c$, the presence of a longitudinal external field $\lambda$
drives FOQTs along the $\lambda=0$ line.  The behavior along the FOQT line
for $g<g_c$ is related to the level crossing of the two lowest states
$|+\rangle$ and $|-\rangle$ for $\lambda=0$, such that $\langle+|
\sigma_{\bf x}^{(3)} |+\rangle = m_0$ and $\langle-|\sigma_{\bf
  x}^{(3)}|-\rangle = -m_0$ (irrespective of ${\bf x}$), with $m_0>0$.
The degeneracy of these states is lifted by the longitudinal field
$\lambda$. Therefore, $\lambda = 0$ is a FOQT point, where the
longitudinal magnetization $M = L^{-d} \sum_{\bf x} M_{\bf x}$, with
$M_{\bf x}\equiv \langle \sigma_{\bf x}^{(3)} \rangle$, becomes
discontinuous in the infinite-volume limit.  The FOQT separates two
different phases characterized by opposite values of the magnetization
$m_0$, i.e.  $\lim_{\lambda \to 0^\pm} \lim_{L\to\infty} M = \pm m_0$.
For one-dimensional systems~\cite{Pfeuty-70}, $m_0 = (1 - g^2)^{1/8}$.

In a finite system of size $L$, the two lowest states are
superpositions of two magnetized states $| + \rangle$ and $| -
\rangle$, in particular when the boundary conditions are neutral, i.e,
they do not favor any of the two magnetized phases, such as periodic
and open boundary conditions (see e.g. Refs.~\cite{CNPV-14, CPV-15,
  PRV-18c} for discussions of the scenario emerging when boundary
conditions are not neutral).  Due to tunneling effects, the energy gap
$\Delta$ at $\lambda=0$ vanishes exponentially as $L$
increases,~\cite{PF-83, CNPV-14}
\begin{equation}
\Delta(L) \sim e^{-c L^d},
\label{expbede}
\end{equation}
apart from powers of $L$.  In particular, the energy gap $\Delta(L)$
of the one-dimensional Ising ring~(\ref{hisdef}) for $g<1$ is
exponentially suppressed as~\cite{Pfeuty-70, CJ-87}
\begin{subequations}
  \begin{equation}
    \Delta(L) = 2 \, (1-g^2) g^L \, [1+ O(g^{2L})]
    \label{deltaobc}
  \end{equation}
for open boundary conditions, and
  \begin{equation}
    \Delta(L) \approx 2 \,\sqrt{(1-g^2)/(\pi L)} \, g^L
    \label{deltapbc}
  \end{equation}
\end{subequations}
for periodic boundary conditions.  The differences for the higher excited
states are finite, for $L\to \infty$.

In the following, we mostly consider quench protocols where the
transverse field $g$ is kept fixed, and $\lambda$ is varied according
to the various quench protocols described above. At $g=g_c$, the
quench protocol provides a paradigmatic example of a CQT, while when
$g<g_c$ it provides examples of FOQTs.

\subsection{Work fluctuations at the continuous transition}
\label{sec:Ising_CQT}

We are now ready to present a numerical verification of the predicted
DFSS behaviors for the work distribution associated with a quench of
the one-dimensional quantum Ising model in the longitudinal field, at
the continuous transition (see Sec.~\ref{fsscqt}).  In particular, we
shall consider chains of size $L$ with periodic boundary conditions,
fix $g=1$ at the critical point, and study zero-temperature quench
protocols driven by the longitudinal field, $\lambda_0 \to \lambda$,
cf. Eq.~(\ref{hisdef}).

\begin{figure}[!t]
  \begin{center}
  \includegraphics[width=0.8\columnwidth]{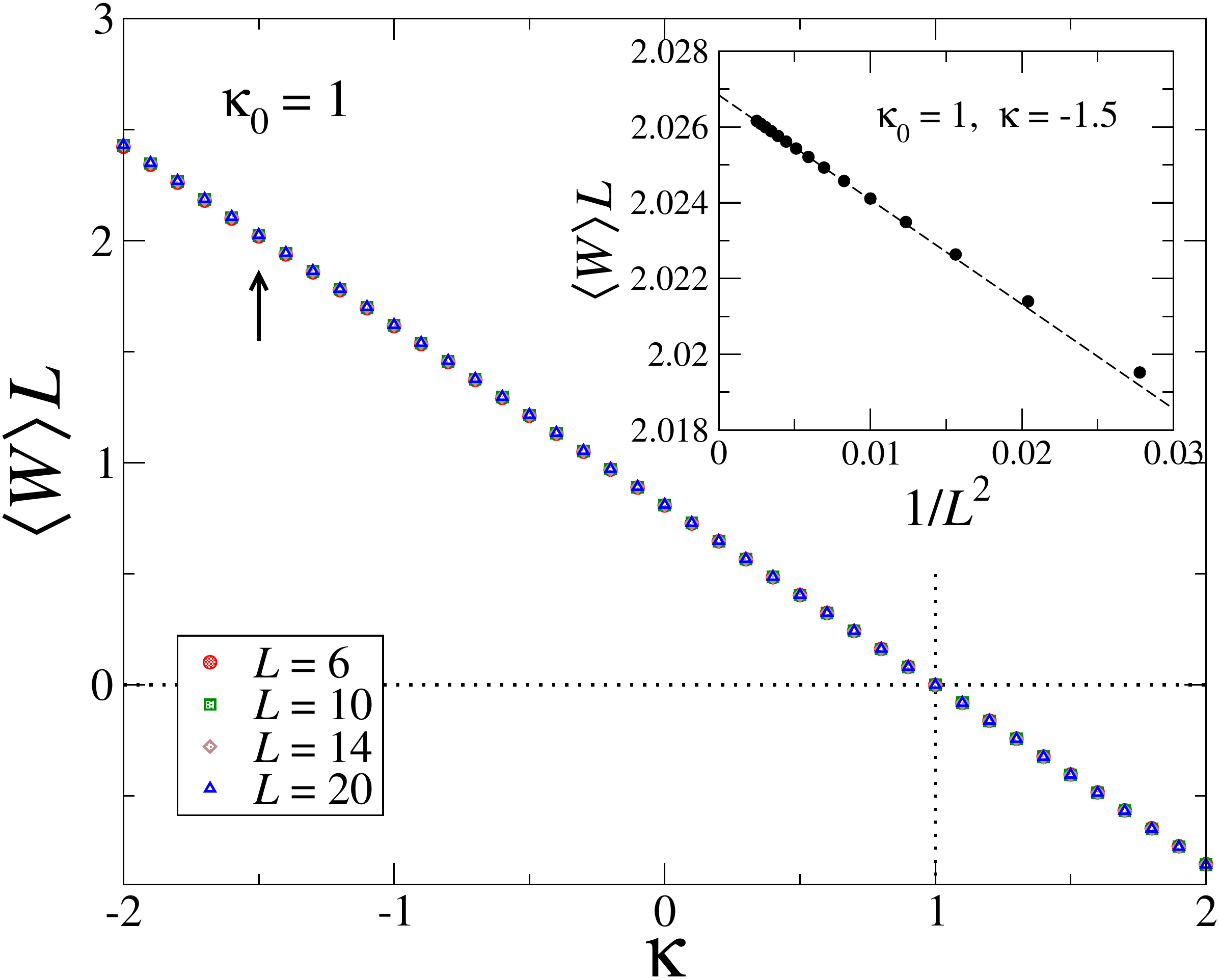}
  \caption{Average work $\langle W \rangle$, rescaled by a factor
    $L^z=L$, injected into the system after a quench of the Ising ring
    at the CQT, for fixed $\kappa_0=1$ and varying $\kappa$.  Notice
    that $\kappa = \kappa_0$ corresponds to the equilibrium point, in
    which no work is performed on the system.  The various data sets
    correspond to different chain lengths, as indicated in the legend.
    The inset shows the behavior of the same data with the system
    size, for fixed $\kappa = -1.5$ (arrow in the main frame).  The
    data appear to converge with $O(L^{-2})$ corrections.  }
  \label{Work1_CQT}
  \end{center}
\end{figure}

Numerical diagonalization data concerning the scaling of the average
post-quench work $\langle W \rangle$, defined in Eq.~(\ref{energy}),
are provided in Fig.~\ref{Work1_CQT}, for a quench starting from
$\kappa_0 = 1$ and for different values of $L$ up to 20
spins~\cite{Note}.  The average work is plotted against the
post-quench renormalized field $\kappa$.  We recall that, for the
model under investigation, the RG dimension of the $\lambda$ parameter
is $y_\lambda = 15/8$, therefore Eq.~(\ref{eq:KappaCQT}) implies that
$\lambda = L^{-15/8} \kappa$ asymptotically approaches zero, in the
large-$L$ limit (and analogously for $\lambda_0$).

At a first glance, one realizes that data collapse emerges quite
neatly already at very small sizes, on the scale of the figure.  This
obeys the scaling predicted by Eq.~(\ref{wscacot0}), after noticing
that $\Delta(L) \sim L^{-z}$, and that $z=1$ for the one-dimensional
quantum Ising ring.  Therefore, the observed linear behavior of
$\langle W \rangle L$ as a function of $\kappa$ (for fixed $\kappa_0$)
immediately follows from the scaling prediction, together with the
identification of ${\cal W}_1$ with that in Eq.~(\ref{w1co}).
We have also analyzed finite-size corrections: the inset displays
numerical data for a cut of the main frame at $\kappa=-1.5$.  The
approach to the asymptotic behavior appears to be characterized by
$O(L^{-2})$ corrections.

\begin{figure}[!t]
  \begin{center}
  \includegraphics[width=0.8\columnwidth]{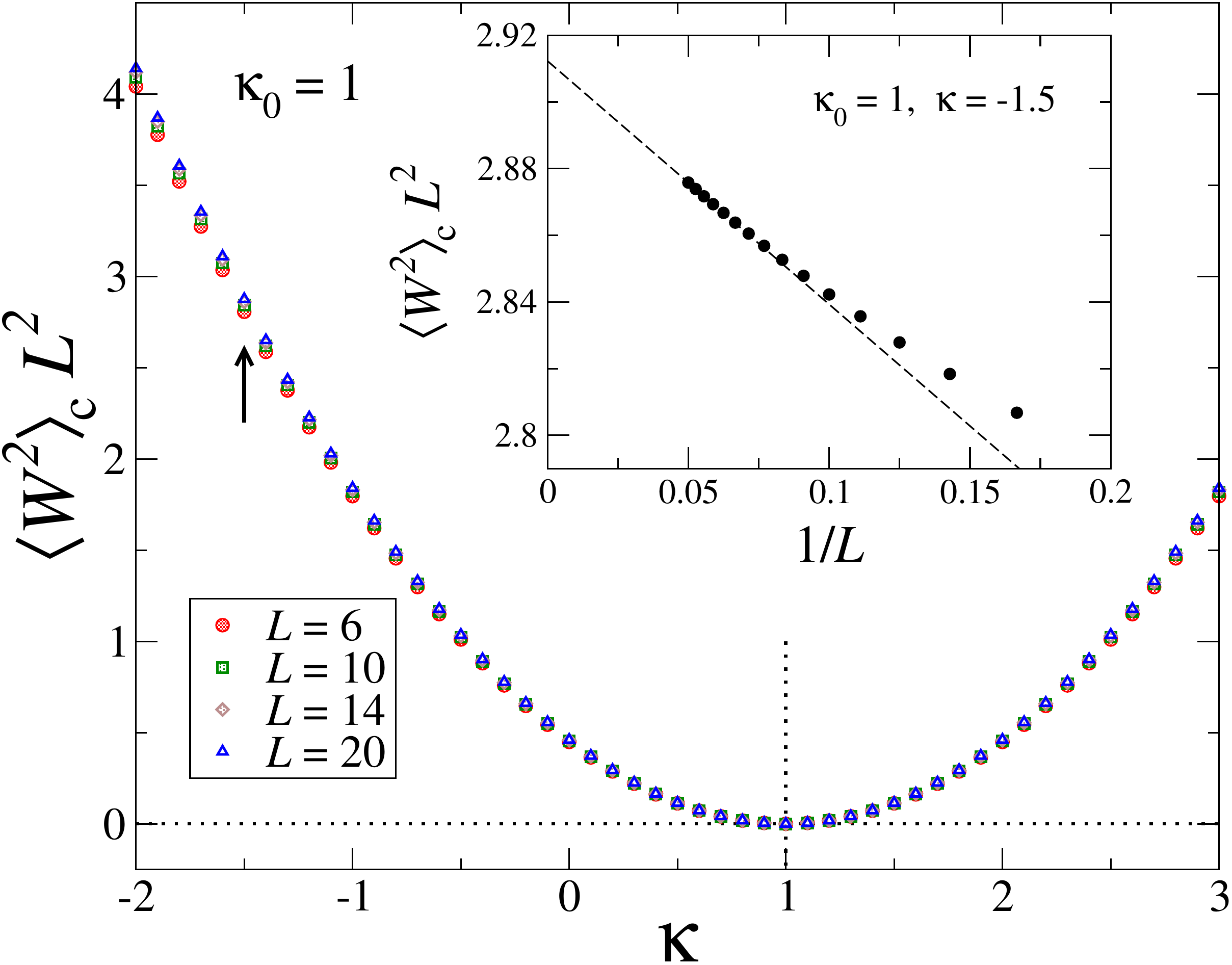}
  \caption{Same as in Fig.~\ref{Work1_CQT}, but for the variance of
    the work distribution, $\langle W^2 \rangle_c = \langle W^2
    \rangle - \langle W \rangle^2$, injected after the quench.  The
    inset highlights the convergence to the asymptotic scaling
    behavior, which appears to follow a $O(L^{-1})$ dependence.
    Unspecified parameters have been chosen to be
    the same as in Fig.~\ref{Work1_CQT}.}
  \label{Work2_CQT}
  \end{center}
\end{figure}

A closer inspection of higher moments of the work distribution is
useful to test the scaling Ansatz put forward in Eq.~(\ref{scalwk}).
To this purpose, we have analyzed the numerical data for the second
moment $\langle W^2\rangle$ of the work distribution, as well.
Figure~\ref{Work2_CQT} displays the connected correlation function
\begin{equation}
\langle W^2 \rangle_c \equiv \langle W^2 \rangle - \langle W
\rangle^2
\label{cw}
\end{equation}
suitably rescaled by $L^{2z}=L^2$, as a function of the scaling
variable $\kappa$, for the same set of parameters as in
Fig.~\ref{Work1_CQT}.  Even in this case, we observe a remarkable
agreement with the predicted DFSS behavior.  We have checked that the
predicted scaling occurs for several other choices of the system
parameters, namely by varying $\kappa_0$ and $\kappa$ (not shown).
The approach to the expected asymptotic behavior of the variance
$\langle W^2 \rangle_c$ turns out to be slower than the average work
$\langle W \rangle$, as is visible in the inset (see the scale on the
$y$-axis).  Indeed, corrections appear to get suppressed as
$O(L^{-1})$. This may suggest that the global convergence of the DFSS
of the work statistics may be $O(L^{-1})$, similarly to the DFSS of
the bipartite entanglement entropy~\cite{PRV-18b}.  We believe that
this issue deserves further investigation.

It is significant to remark that the statistics of the work in the
DFSS limit is generally not Gaussian, therefore higher moments of the
work distribution are also important to be analyzed.  In view of the
fact that an increasing amount of computational resources is required
to obtain $\langle W^k \rangle$, with $k>2$, we decided to directly
tackle the characteristic function of Eq~(\ref{csamp}) at finite
times.  To this purpose, instead of fully diagonalizing the
post-quench Hamiltonian $H(\lambda)$, we have implemented the time
evolution of the initial state $|0_{\lambda_0}\rangle$ by means of a
fourth-order Suzuki-Trotter decomposition of the unitary-evolution
operator $U(t) = e^{-i H(\lambda) t}$, with a time step $\delta
t=10^{-2}$, for systems with up to $L=23$ sites.

\begin{figure}[!t]
  \begin{center}
  \includegraphics[width=0.8\columnwidth]{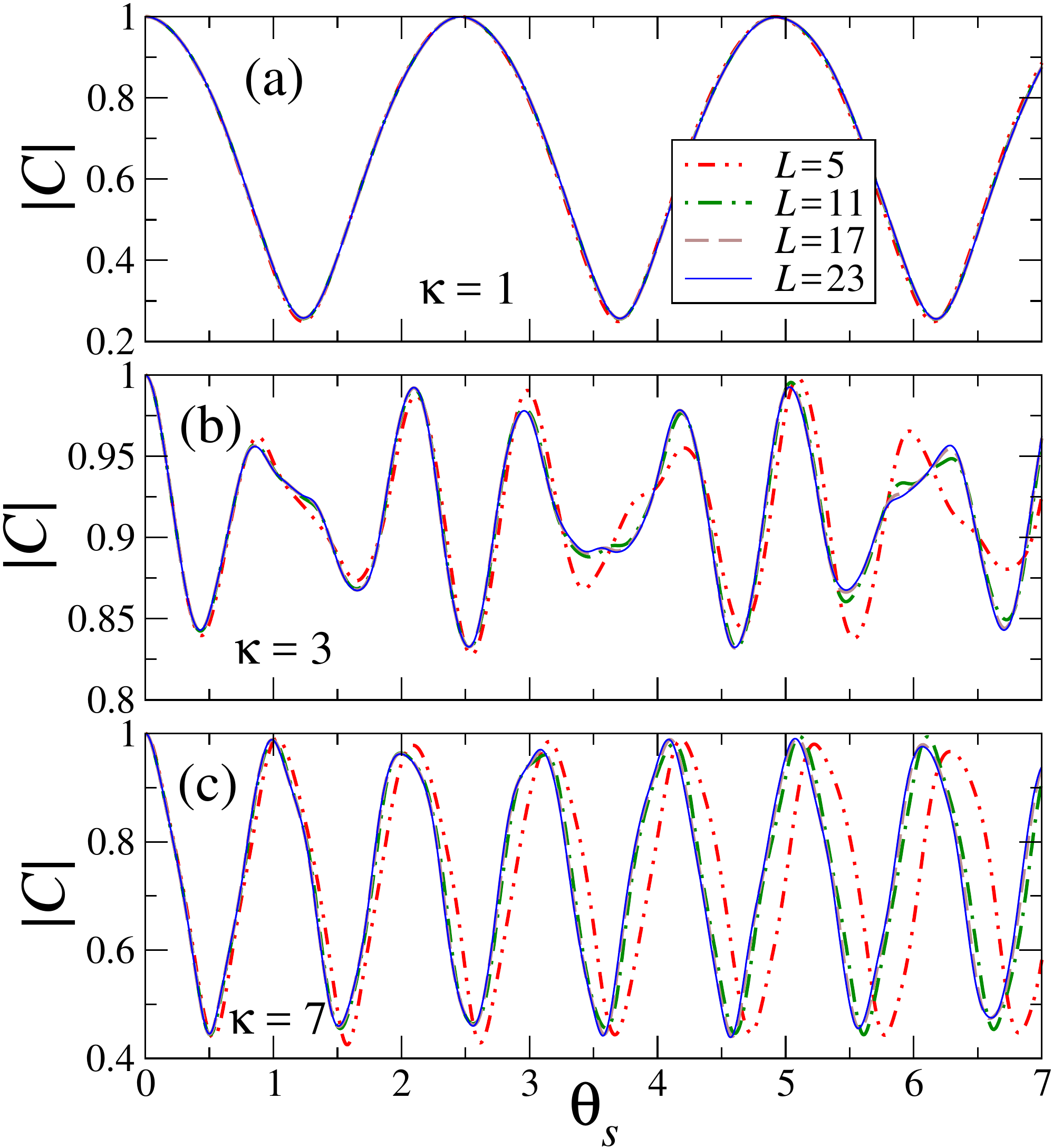}
  \caption{Scaling of the modulus of the characteristic function
    $|C(s)|$ associated to a quench of the Ising model for $\kappa =
    -\kappa_0$. From top to bottom, the three panels refer to
    $\kappa=1$, $3$, and $7$, respectively.  Data are plotted against
    the rescaled time $\theta_s$.  Notice that, for the sake of
    convenience, in this figure we have defined $\theta_s = L^z \, s$,
    cf. Eq.~(\ref{thetadef}).}
  \label{CharFun_CQT}
  \end{center}
\end{figure}

We have eventually checked the DFSS prediction of Eq.~(\ref{gtsca}),
which entails the validity of the Ansatz for the whole distribution of
the work statistics.  The corresponding plot for the modulus of
$C(s)$, as a function of the rescaled time variable $\theta_s$, is
reported in Fig.~\ref{CharFun_CQT} for three different values of
$\kappa_0$, with $\kappa = -\kappa_0$.  Even for this quantity, data
collapse is remarkably evident at the relatively small sizes we were
able to reach, with a slightly slower approach to the asymptotic
scaling behavior when increasing $\kappa$.  The convergence
to the asymptotic DFSS behavior of $C(s)$ appears consistent with
a global $s$-dependent $O(1/L)$ suppression of the corrections. 
Notice that the irregular
pattern of the various curves signals the presence of non-Gaussian
features in the statistics of the work.  The irregularity of these
curves, and thus the deviations from Gaussianity, is generally non
monotonic in $\kappa$, since it depends on the degree of
commensurability of the energy injected by the quench with the
spectrum of the system~\cite{PRV-18b}.

Note that the main features of the DFSS at the CQT of the Ising models
(\ref{hisdef}), such as the general size dependence and the scaling
functions, are expected to be universal, i.e., they are expected not
to depend on the microscopic details of the models (apart from trivial
normalizations of the arguments). Therefore, their predictions can be
extended to all CQTs belonging to the corresponding Ising universality
classes.

\subsection{Work fluctuations at the first-order transitions}
\label{dfssfoqt}

Let us now switch to the work distribution associated with quenches
along the FOQT line of the model in Eq.~(\ref{hisdef}), i.e., driven
by the longitudinal external field $\lambda$ along the line $g<g_c$ of
the phase diagram.
As already mentioned, isolated many-body systems at FOQTs develop a
FSS behavior, as well.  However, they significantly depend on the type
of boundary conditions, in particular whether they favor one of the
phases or they are neutral, giving rise to FSS characterized by
exponential or power-law behaviors.  In the following we shall
consider Ising systems with boundary conditions that do not favor any
of the two magnetized phases, such as periodic and open boundary
conditions, which generally lead to exponential FSS
laws~\cite{CNPV-14}.  We stress that, for peculiar boundary conditions
as the antiperiodic ones, the energy difference of the lowest levels
obeys a power-law dependence on $L$, in which case the situation
becomes more subtle~\cite{CNPV-14}.

Following the general DFSS framework put forward in
Sec.~\ref{dfsscqt}, we can identify the scaling parameters $\kappa_0$,
$\kappa$, and $\theta_t$ (see Sec.~\ref{dfsscqt}).  Specifically, as
described in Sec.~\ref{sec:Ising_CQT}, we perform the quench protocol
$\lambda_0 \to \lambda$ for a value of $g<g_c$.  The energy associated
with the corresponding longitudinal-field perturbation $\lambda$ is
given by $E_\lambda(\lambda,L)=2 m_0 \lambda_0 L^d$, while the gap
$\Delta(L)$ of the two lowest states at $\lambda=0$ is given by
Eqs.~(\ref{expbede}), (\ref{deltaobc}), and~(\ref{deltapbc}),
depending on the boundary conditions.  We can thus express the scaling
variables $\kappa_0$ and $\kappa$ as~\cite{PRV-18b}
\begin{equation}
  \kappa_0 = {2 m_0 \lambda_0 L^d\over \Delta(L)},\qquad
  \kappa = {2 m_0 \lambda \,L^d \over \Delta(L)}.
  \label{kappa_FOQT}
\end{equation}
The corresponding DFSS of the work distribution, Eq.~(\ref{scalwk}),
is expected to hold for any $g<g_c$, with a scaling function ${\cal
  P}$ independent of $g$, apart from trivial normalizations of the
arguments. The approach to the asymptotic scaling is expected to be
exponential when increasing the size of the system (for the cases
under consideration, the gap $\Delta$ closes exponentially with $L$).

In the case of the quantum Ising systems with periodic or open
boundary conditions, the DFSS functions can be exactly computed,
exploiting a two-level truncation of the spectrum~\cite{CNPV-14,
  PRV-18, PRV-18b}.  In the long-time limit and for large systems, the
scaling properties in a small interval around $\lambda=0$, more
precisely for $m_0 |\lambda|\ll \Delta_2=O(1)$, are captured by a
two-level truncation, which only takes into account the two nearly
degenerate lowest-energy states.  The effective evolution is
determined by the Schr\"odinger equation
\begin{equation}
i {d\over dt} \Psi(t) = H_{2l}(\lambda) \, \Psi(t) ,
\label{sceq}
\end{equation}
where $\Psi(t)$ is a two-component wave function, whose components
correspond to the states $|+ \rangle$ and $|-\rangle$, and
\begin{equation}
  H_{2l}(\lambda) = - \beta \, \sigma^{(3)} + \delta \, \sigma^{(1)},
  \qquad \mbox{with } \;\; \beta = m_0 \lambda L^d,\quad \delta =
  \Delta(L) / 2.
  \label{hrtds}
\end{equation}
Using Eq.~(\ref{kappa_FOQT}), we also have $\beta/\delta = \kappa$.
The initial condition is given by the ground state of
$H_{2l}(\lambda_0)$:
\begin{equation}
|\Psi(\lambda_0,\lambda,L,t=0)\rangle = \sin(\alpha_0/2) \, |-\rangle -  
\cos(\alpha_0/2) \, |+\rangle, \label{eigstatela0}
\end{equation}
with $\tan \alpha_0 = \kappa_0^{-1}$ and $\alpha_0\in (0,\pi)$.  The
quantum evolution after quenching $\lambda_0 \to \lambda$ can be
easily obtained by diagonalizing $H_{2l}(\lambda)$. Its eigenstates,
associated with the eigenvectors $E_{0,1}^\lambda = \mp \Delta
\sqrt{1+\kappa^2} / 2$, are
\begin{equation}
  |0\rangle = \sin(\alpha/2) \, |-\rangle - \cos(\alpha/2) \, |+\rangle\,, \qquad
  |1\rangle = \cos(\alpha/2) \, |-\rangle + \sin(\alpha/2) \, |+\rangle\,,
  \label{eigstate1la}
\end{equation}
with $\tan \alpha = \kappa^{-1}$ and $\alpha\in (0,\pi)$.  
Then, it is not difficult to show that, after the quench at $t=0$, the
state in Eq.~(\ref{eigstatela0}) evolves as
\begin{equation}
  |\Psi(\lambda_0,\lambda,L,t)\rangle = e^{i \frac{\theta_t}{2}
    \sqrt{1 + \kappa^2}} \cos (\delta \alpha) |0\rangle 
  + e^{-i \frac{\theta_t}{2} \sqrt{1 + \kappa^2}} \sin (\delta \alpha) |1\rangle,
  \label{psitfo}
\end{equation}
where we defined $\delta \alpha = (\alpha_0 - \alpha)/2$.  Note that
the time-dependent wave function in Eq.~(\ref{psitfo}) is written in
terms of scaling variables only.

Using these results, one can easily compute the characteristic
function defined in Eq.~(\ref{csamp}), that is, the amplitude:
\begin{equation}
  C(s) =
  e^{-i E_0 s} \langle \Psi(\lambda_0,\lambda,L,s=0)|
\Psi(\lambda_0,\lambda,L,-s)\rangle.  
\label{csexp}
\end{equation}
Therefore, the work probability distribution $P(W)$ can be inferred
directly from Eq.~(\ref{gudef}).  Its corresponding scaling function
[see Eq.~(\ref{genpwsca})] reads:
\begin{eqnarray}
&&  {\cal P}^{(2l)}(\omega,\kappa_0,\kappa) = 
  \delta(\omega - \omega_-) \cos^2 (\delta \alpha) +
  \delta(\omega - \omega_+) \sin^2 (\delta \alpha), \nonumber
  \\ &&\mbox{with }\;\quad \omega_\pm =  {1\over 2} \Big( \sqrt{1 +
    \kappa_0^2} \pm \sqrt{1 + \kappa^2} \Big).
\end{eqnarray}
This is in agreement with the expected scaling behavior.

Let us now specialize to the average work $\langle W \rangle =
\Delta(L) \,{\cal W}_1^{(2l)}(\kappa_0,\kappa)$.  After simple
manipulations, working in the hypothesis of a two-level $(2l)$
truncation of the spectrum, this turns out to be characterized by the
scaling function
\begin{equation}
{\cal W}_1^{(2l)} (\kappa_0,\kappa) = 
{(\kappa_0 - \kappa)\kappa_0 \over 2\sqrt{1+\kappa_0^2}}. \label{pred_w1}
\end{equation}
This expression has the same form of Eq.~(\ref{w1co}).
It is instructive to see that it satisfies the inequality
\begin{equation}
  {\cal W}_1^{(2l)}  >  {1\over 2} \Big( \sqrt{1 + \kappa_0^2}
  - \sqrt{1 + \kappa^2} \Big),
  \label{inavw}
\end{equation}
or, equivalently,
\begin{equation}
  \langle W \rangle > E^\lambda_0 - E^{\lambda_0}_0.
\end{equation}

We remark that in the DFSS limit, the effects of higher states are
expected to be exponentially suppressed, essentially because in this
limit the probability associated with higher states is exponentially
suppressed~\cite{PRV-18}.

\begin{figure}[!t]
  \begin{center}
  \includegraphics[width=0.8\columnwidth]{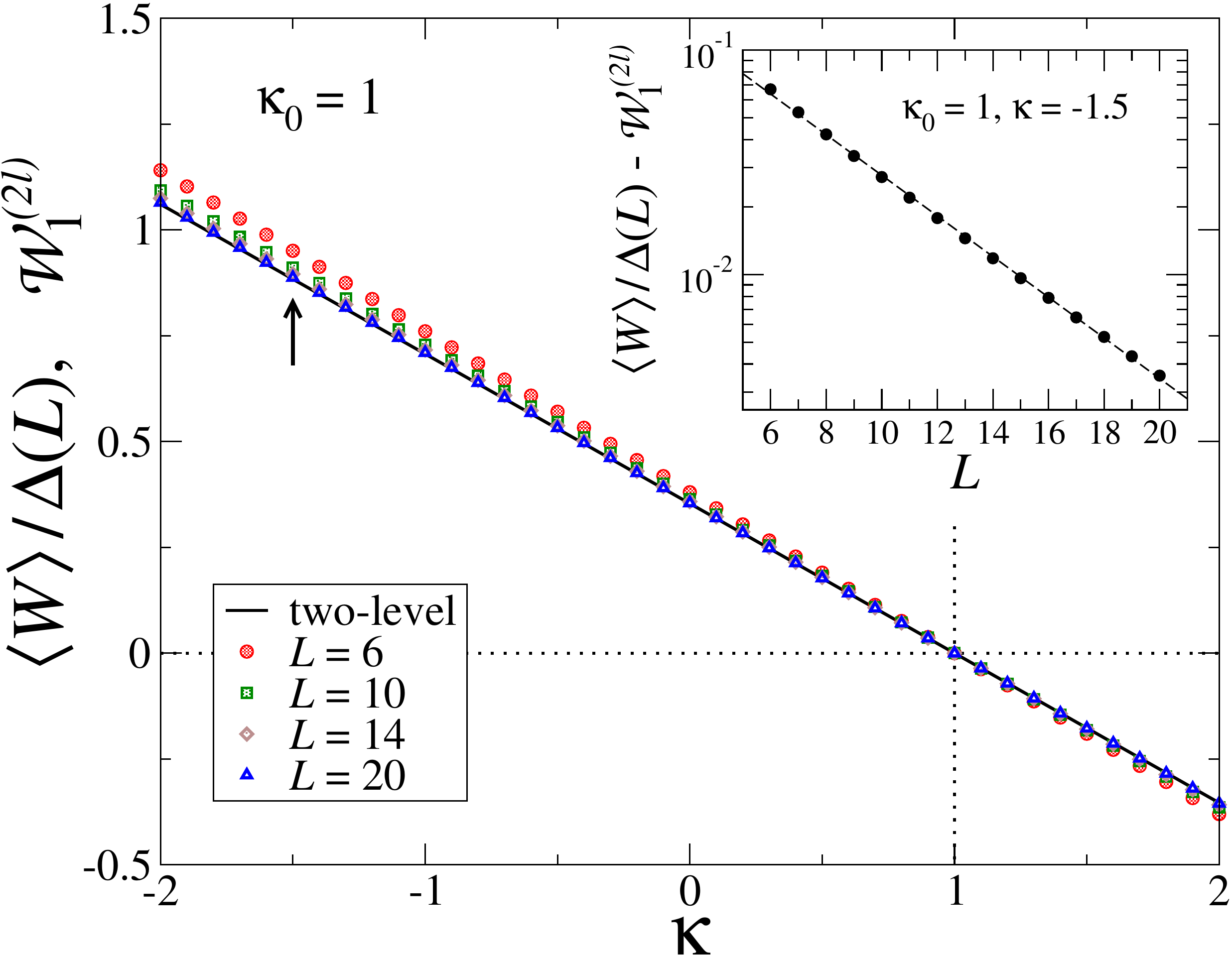}
  \caption{Average work $\langle W \rangle$, divided by the
    ground-state energy gap $\Delta(L)$, after a quench across the
    FOQT of the Ising ring with $g=0.9$ and $\kappa_0=1$, as a
    function of $\kappa$ and for different chain lengths.  The
    continuous straight line is the two-level prediction of
    Eq.~(\ref{pred_w1}).  The inset highlights the difference between
    the numerical results at finite size and the analytic function
    ${\cal W}_1^{(2l)}$, which appears to decrease exponentially
    (the dashed line is an exponential fit of the data).}
  \label{Work1_FOQT}
  \end{center}
\end{figure}

To corroborate the above DFSS predictions in the two-level truncation
hypothesis close to a FOQT, we have numerically computed the average
work after a quench in the longitudinal field $\lambda$ for the
one-dimensional quantum Ising chain at finite size $L$, with $g<1$.
The corresponding data for the ratio $\langle W \rangle / \Delta(L)$,
at fixed $g=0.9$ and $\kappa_0=1$, are shown in Fig.~\ref{Work1_FOQT}
as a function of the rescaled variable $\kappa$.  A comparison of the
numerical outcomes with the analytic estimate ${\cal W}_1^{(2l)}$
of Eq.~(\ref{pred_w1}) demonstrates a remarkable agreement between
them, as is visible from the figure.  A closer inspection to
the behavior with $L$, for fixed $\kappa$, is provided in the inset,
where an exponential convergence to the expected prediction clearly
emerges.  An analogous scaling behavior occurs for other values of
$g<1$ (not shown), thus showing universality within the Ising-like
FOQTs.

Let us close this discussion by mentioning that the above derived
zero-temperature scaling formulas can be easily extended to finite
temperature as well, using Eq.~(\ref{pwdefft}).  This requires the
further scaling variable $\tau = \Delta(L)^{-1} \,T$. The
corresponding work distribution satisfies the Jarzynski equality, see
Eq.~(\ref{jeq}):
\begin{equation}
  \int dW e^{-\beta W} P(W,T,\lambda_0,\lambda,L) =
  e^{-\beta[F(\lambda)-F(\lambda_0)]} .
  \label{jaeequ}
\end{equation}
More precisely, using the fact that 
\begin{equation}
P(W,T,\lambda_0,\lambda,L) =
\Delta(L)^{-1} \, {\cal P}^{(2l)}(\omega,\tau,\kappa_0,\kappa),
\label{2lpw}
\end{equation}
cf. Eq.~(\ref{genpwsca}), we obtain
\begin{equation}
  \int d\omega\, e^{-\omega/\tau}{\cal P}^{(2l)}
  (\omega,\tau,\kappa_0,\kappa) = { {\rm cosh} \big[
      \sqrt{1+\kappa^2}/ (2\tau) \big]\over {\rm cosh} \big[
      \sqrt{1+\kappa_0^2} / (2\tau) \big] }.
  \label{jarsca}
\end{equation}

We finally address the quench protocol introduced at the end of
Sec.~\ref{qhprwd}. The quantum evolution starts from the ground state
associated with the parameter value $\lambda_0$, which is suddenly
changed to $\lambda$ at $t=0$. After a time $t$ the parameter is
quenched back to the value $\lambda_0$.  In the case of FOQTs,
two-level computations analogous to those employed for the standard
quench protocol confirm the conjectured scaling behaviors and allow us
determine the corresponding scaling functions.  We obtain
\begin{equation}
{\cal P}^{(2l)}(\omega,\kappa_0,\kappa,\theta_t)= A \, \delta(\omega)
+ (1-A) \: \delta \Big(\omega - \sqrt{1+\kappa_0^2} \Big),
\end{equation}
where $A \equiv |\langle 0_{\lambda_0}|\Psi(t)\rangle|^2$ denotes the
overlap between the initial state and the time-evolved state,
cf. Eq.~(\ref{psitfo}), and is given by
\begin{equation}
A(\kappa_0,\kappa,\theta_t) = 1 - \frac12 \Big[1 - \cos(\theta_t
  \sqrt{1+\kappa^2}) \Big] \sin^2(\alpha_0 - \alpha).
 \label{scalpiwidq}
\end{equation}

Finally, we would like to stress that the above DFSS of the work
fluctuations is expected to be the same, apart from normalizations,
along the FOQT line of the quantum Ising models (\ref{hisdef}) for
$g<g_c$, and in any system sharing the same global properties, such as
FOQTs arising from an avoided two-level crossing phenomenon in the
large-$L$ limit.

\section{Results for the Bose-Hubbard model}
\label{xxzmod}

\subsection{The lattice Bose-Hubbard model}
\label{BHmodel}

Another physically interesting system is the Bose-Hubbard (BH)
model~\cite{FWGF-89}, which provides a realistic description of a gas
of bosonic atoms in an optical lattice~\cite{JBCGZ-98}. Its
Hamiltonian reads:
\begin{equation}
H_{\rm BH} = - {J\over 2} \sum_{\langle {\bf x},{\bf y}\rangle} 
(b_{\bf x}^\dagger b_{\bf y} + b_{\bf y}^\dagger b_{\bf x})
+{U\over 2} \sum_{\bf x} n_{\bf x}(n_{\bf x}-1) 
-\mu \sum_{\bf x} n_{\bf x} ,
\label{bhm}
\end{equation}
where $b_{\bf x}$ annihilates a boson on site ${\bf x}$ of a cubic
$L^d$ lattice, $n_{\bf x}\equiv b_{\bf x}^\dagger b_{\bf x}$ is the
particle density operator, the first sum runs over the
nearest-neighbor bonds ${\langle {\bf x},{\bf y}\rangle}$, while the
others run over the sites.  Moreover the parameter $J$ denotes the
hopping strength, $U$ the interaction strength, and $\mu$ the onsite
chemical potential. In addition, we consider a perturbation coupled to
the particle operator $b_{\bf x}$, i.e.,
\begin{equation}
  H(\lambda) = H_{\rm BH} + \lambda  H_p, \,,\qquad
  H_p = - {1\over 2} \sum_{\bf x} (b_{\bf x} + b_{\bf x}^\dagger),
\label{hlabh}
\end{equation}
where the parameter $\lambda$ plays the role of a real external
constant field.  We express lengths in terms of the lattice spacing
$a=1$ and set $J=1$, so that energies are provided in units of $J$.

Here we are interested in the infinitely repulsive, hard-core
$U\to+\infty$ limit, so that the particle number can only take the
values $n_{\bf x}=0,\,1$.  In this case, the BH Hamiltonian
$H(\lambda)=H_{\rm BH}+\lambda H_p$, cf. Eq.~(\ref{hlabh}), can be
exactly mapped into the so-called XX model~\cite{Sachdev-book}
\begin{equation}
  H_{\rm XX} = - \sum_{\langle {\bf x},{\bf y}\rangle} \left[
    S^{(1)}_{\bf x}S^{(1)}_{\bf y} + S^{(2)}_{\bf x}S^{(2)}_{\bf
      y}\right] + \mu \sum_{\bf x} \Big( S^{(3)}_{\bf
    x}-\frac12 \Big) - \lambda \sum_{\bf x} S^{(1)}_{\bf x},
  \label{XXH}
\end{equation}
where the spin operators $S_{\bf x}^{(k)}=\sigma_{\bf x}^{(k)}/2$ are
related to the bosonic ones by: $\sigma_{\bf x}^{(1)} = b_{\bf
  x}^\dagger + b_{\bf x}$, $\sigma^{(2)}_{\bf x} = i(b_{\bf x}^\dagger
- b_{\bf x})$, and $\sigma^{(3)}_{\bf x} = 1-2b_{\bf x}^\dagger b_{\bf
  x}$.

The zero-temperature limit of the hard-core BH model, or equivalently
of the XX model, displays three phases for $\lambda=0$, associated
with the ground-state properties~\cite{Sachdev-book}: the vacuum ($\mu
<-d$), the superfluid ($- d < \mu < d$), and the Mott $n=1$ phase
($\mu > d$).  The vacuum-to-superfluid transition at $\mu_{\rm vs}=-d$
and the $n=1$ superfluid-to-Mott transition at $\mu_{\rm sm}=d$, when
they are driven by the chemical potential, belong to the universality
class associated with a {\em nonrelativistic} $U(1)$-symmetric bosonic
field theory~\cite{FWGF-89}.  The upper critical dimension of this
bosonic field theory is $d=2$. Thus its critical behavior is mean
field for $d>2$.  For $d=2$ the field theory is essentially free
(apart from logarithmic corrections), thus the dynamic critical
exponent is $z=2$, the RG dimension of the coupling $\mu$ is $y_\mu =
2$.  In $d=1$ the theory turns out to be equivalent to a free-field
theory of nonrelativistic spinless fermions~\cite{Sachdev-book}, from
which one infers the RG exponents $z=2$ and $y_\mu=2$.  The RG
dimension of the $\lambda$ parameter of the Hamiltonians~(\ref{hlabh})
and~(\ref{XXH}) is $y_\lambda=z+d/2=2+d/2$ (thus $y_\lambda = 5/2$ for
$d=1$, and $y_\lambda = 3$ for $d=2$).

\subsection{Work fluctuations in quenches 
at the vacuum-to-superfluid transition}
\label{numres}

Let us concentrate on quench protocols at the vacuum-to-superfluid
transition of the hard-core BH model~(\ref{hlabh}), driven by the
Hamiltonian variable $\lambda$, around the critical point $\mu_{\rm
  vs}=-d$ and $\lambda=0$.  According to the general DFSS theory
outlined in Sec.~\ref{dfsswd}, in particular Sec.~\ref{fsscqt}, we
expect the scaling behavior for the work distribution reported in
Eq.~(\ref{Psca}), with $z=2$ and $y_\lambda=2+d/2$.

\begin{figure}[!t]
  \begin{center}
  \includegraphics[width=0.8\columnwidth]{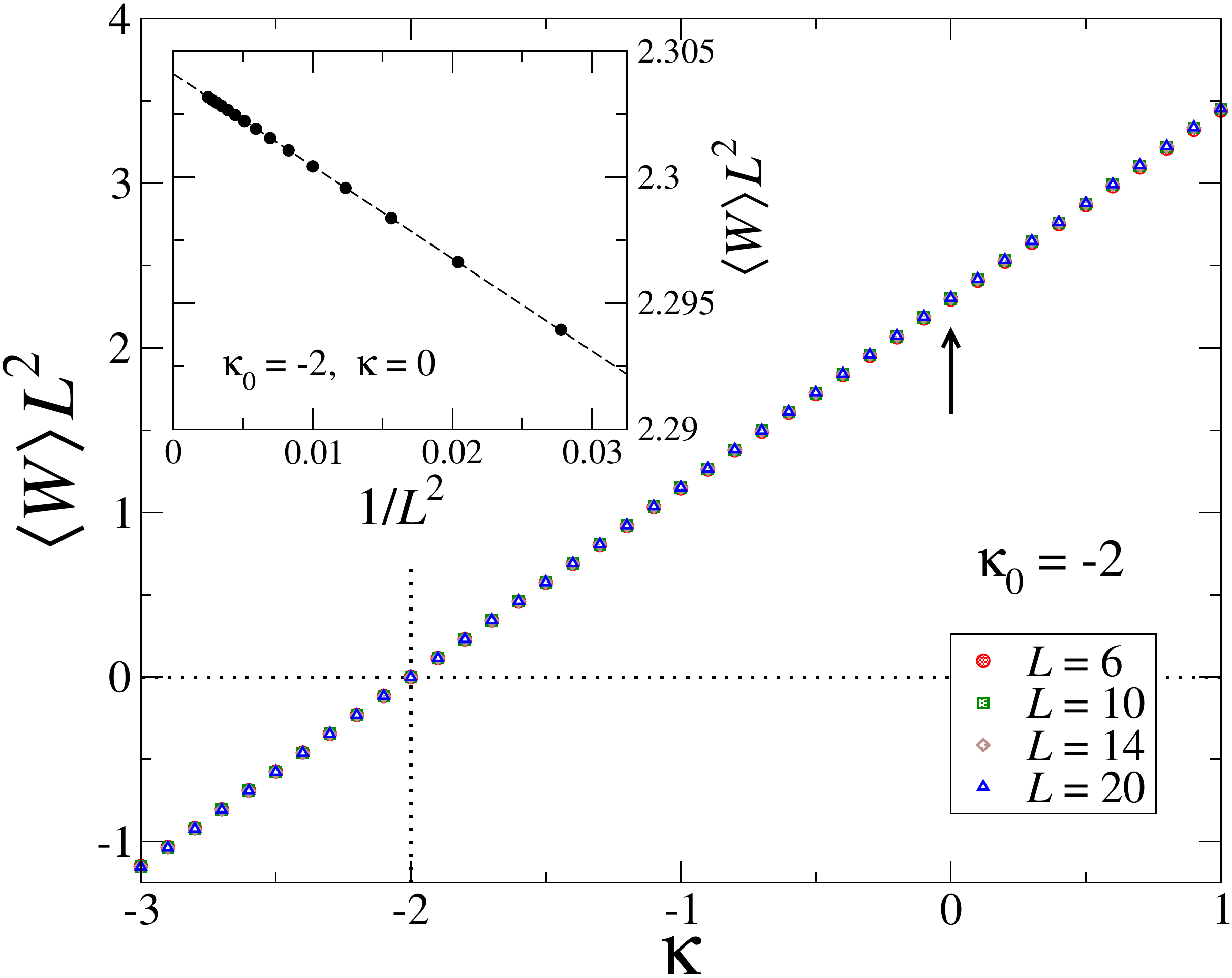}
  \caption{Average work $\langle W \rangle$ for a quench of the XX
    model, rescaled by $L^z = L^2$, at fixed $\kappa_0=-2$ and for
    varying $\kappa$.  The various data sets are for different chain
    lengths.  The inset displays the behavior of the same data with
    the inverse square of the system size, for fixed $\kappa=0$ (arrow
    in the main frame).  They appear to decrease as $1/L^2$.}
  \label{Work1_BH}
    \end{center}
\end{figure}

We have numerically checked these predictions by diagonalizing the XX
chain with periodic boundary conditions~\cite{Note}. The quench
protocol $\lambda_0 \to \lambda$ is performed at zero temperature,
keeping $\mu = \mu_{\rm vs}=-1$ fixed.  Data for the scaling of the
average work $\langle W \rangle$ at a given value of $\kappa_0$ and
for varying $\kappa$ are presented in Fig.~\ref{Work1_BH}.
Analogously to the continuous transition in the quantum Ising chain,
reported in Fig.~\ref{Work1_CQT}, we immediately realize that the data
nicely collapse already for small sizes, exhibiting a linear behavior
of $\langle W \rangle L^z$ with $z=2$ as a function of $\kappa$; this
follows from the scaling prediction in Eqs.~(\ref{wscacot0})
and~(\ref{w1co}).  The convergence to the asymptotic DFSS behavior is
shown in the inset of Fig.~\ref{Work1_CQT}.

\begin{figure}[!t]
  \begin{center}
  \includegraphics[width=0.8\columnwidth]{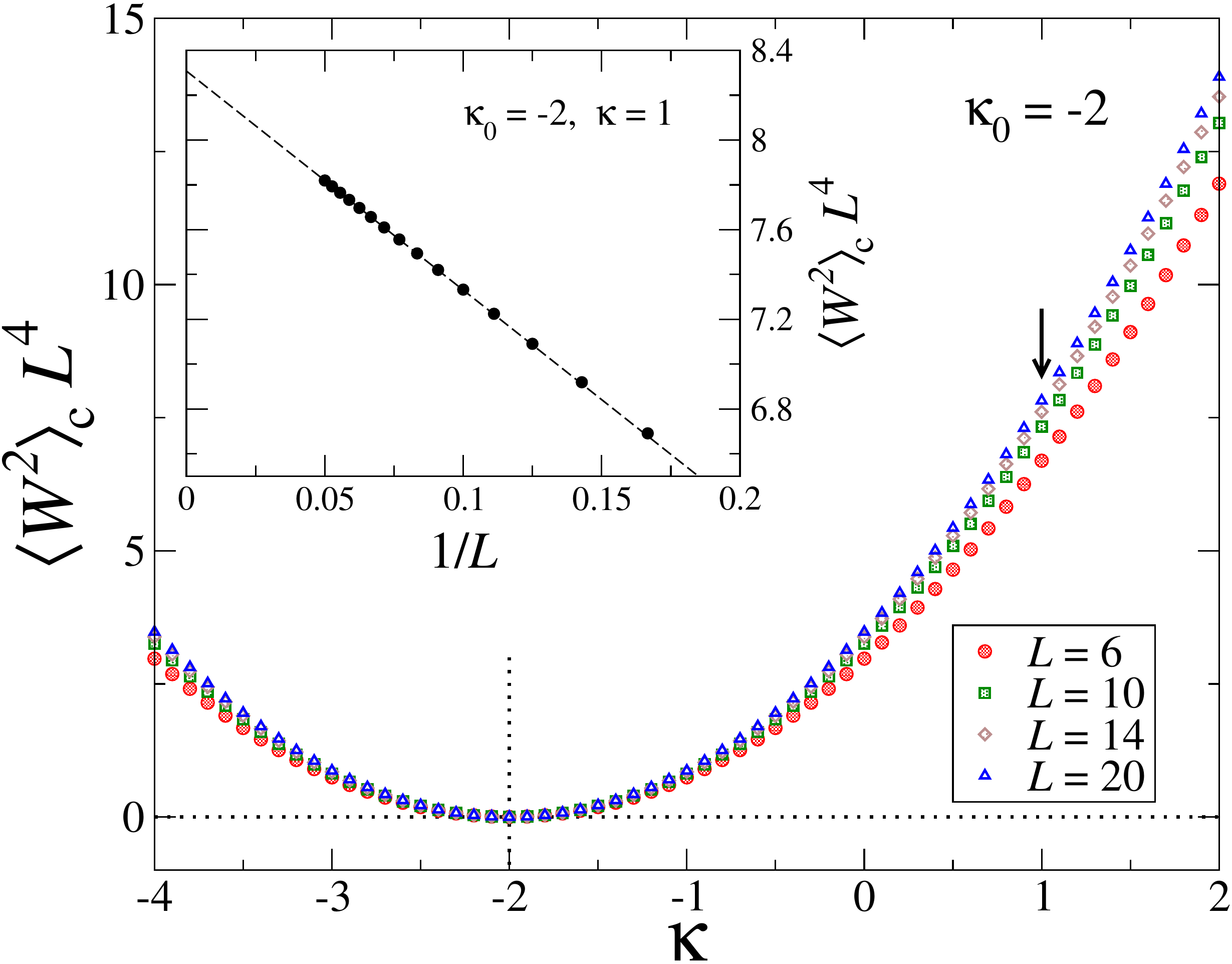}
  \caption{Same as in Fig.~\ref{Work1_BH}, but for the connected
    correlator $\langle W^2 \rangle_c$, rescaled by $L^{2z}=L^4$.  In
    the inset we analyze the behavior with $L$ at fixed $\kappa=1$
    (arrow in the main frame), highlighting the convergence to the
    asymptotic behavior, which appears to be characterized by $O(L^{-1})$
    corrections, similarly to the results for the Ising chain at the
    critical point. Unspecified parameters have been chosen to be
    the same as in Fig.~\ref{Work1_BH}.}
  \label{Work2_BH}
  \end{center}
\end{figure}

The DFSS framework holds for all the moments of the work distribution.
As a matter of fact, we have also analyzed in detail the connected
correlator $\langle W^2\rangle_c$ and checked its scaling properties.
Data collapse can be observed after a suitable rescaling, according to
Eq.~(\ref{scalwk}): pertinent data are presented in
Fig.~\ref{Work2_BH}, where $\langle W^2\rangle_c \, L^{2z}$ is plotted
against $\kappa$.  Analogous results have been found for several other
values of $\kappa_0$ and $\kappa$.

Our results have been obtained in the hard-core $U\to\infty$ limit.
However, the DFSS behavior at the vacuum-to-superfluid transition is
expected to be universal, thus independent of the on-site interaction
strength $U$, apart from trivial normalizations (of course the location
$\mu_{\rm vs}$ does depend on $U$).

We finally point out that one may also consider quench protocols
driven by the Hamiltonian variable $\mu$ around the critical point
$\mu=\mu_{\rm vs}$ and $\lambda=0$.  In such case, unlike the quench
protocol driven by the $\lambda$ parameter, the driving Hamiltonian
term commutes with the rest of the Hamiltonian.  Dynamic scaling
arguments apply as well.  However, to keep this presentation self
contained, we do not pursue this issue further.


\subsection{Work fluctuations arising from variations of confining potentials}
\label{tsswf}

A common feature of any realistic experiment with cold
atoms~\cite{BDZ-08} is the presence of an external (typically
harmonic) potential $V({\bf x})$ coupled to the particle density,
which traps the particles within a limited space region.  This section
is devoted to a generalization of our DFSS considerations on the
statistics of the work in the BH model, to the case in which the
system is confined by a trapping potential.

Let us fix our setting by considering rotationally-invariant power-law
potentials, of the form
\begin{equation}
V({\bf x},\ell) = v^p |{\bf x}|^p \equiv (|{\bf x}|/\ell)^p,
\label{potential}
\end{equation}
where $v$ is the distance from the center of the trap, which we locate
at the origin (${\bf x}=0$) of the axis, $p$ are positive constants
and $\ell\equiv 1/v$ is the trap size~\cite{CV-10-2,RM-04}.
Experiments are usually set up with a harmonic potential, i.e., $p=2$.
In the case of harmonic traps, $\ell\sim \omega^{-1}$ where $\omega$
is the trap frequency.  This trapping force gives rise to a further
inhomogeneous term to be added to the BH Hamiltonian (\ref{bhm}), i.e,
\begin{equation}
H_{\rm tBH} = H_{\rm BH} + \sum_{\bf x} V({\bf x},\ell) \, n_{\bf
  x}. \label{bhmt}
\end{equation}
Far from the origin, the potential $V({\bf x},\ell)$ diverges,
therefore $\langle n_{\bf x}\rangle$ vanishes and the particles are
trapped.

We want to infer the scaling properties of the work fluctuations at
the vacuum-to-superfluid transition, arising from quench protocols
varying the trap frequency, and therefore the trap size.  Namely, we
assume that the system is initially prepared in the ground state
corresponding to the trap size $\ell_0$, and then the trap frequency
is suddenly changed to a different value corresponding to a new trap
size $\ell$, leading to a out-of-equilibrium quantum dynamics.
Similar protocols have been discussed in Refs.~\cite{CV-10-4, V-12,
  CSC-13}.

The inhomogeneity due to the trapping potential strongly affects the
phenomenology of quantum transitions in homogeneous systems.  The
trapping potential~(\ref{potential}) coupled to the particle density,
as in Eq.~(\ref{bhmt}), significantly affects the critical modes,
introducing another {\em trap} length scale $\ell$.  However,
universal behaviors are recovered in the so-called trap-size scaling
(TSS) limit~\cite{CV-09}.  The corresponding TSS behavior describes
the distortion of the critical behavior around the center of the trap,
due to the spatial inhomogeneity arising from the confining potential.
TSS has some analogies with the FSS theory for homogeneous systems,
with two main differences: the inhomogeneity induced by the
space-dependence of the external field, and a nontrivial power-law
dependence of the correlation length $\xi_\ell$ when increasing the
trap size $\ell$ at the critical point, i.e.,
\begin{equation}
\xi_\ell\sim \ell^\theta,
\label{xildef}
\end{equation}
where $\theta$ denotes the universal {\em trap} exponent.  The latter
can be inferred by a RG analysis of the perturbation induced by the
external trapping potential coupled to the particle
density~\cite{CV-09}.  In the case of one-dimensional and
two-dimensional BH models at their vacuum-to-superfluid and Mott
transition driven by the chemical potential, the universal trap
exponent is given by~\cite{CV-10-2}
\begin{equation}
\theta = {p\over p + y_\mu} = {p\over p + 2},
\label{theta}
\end{equation}
thus $\theta=1/2$ for harmonic confining potentials.  Correspondingly,
the gap at the transition points is expected to scale
as~\cite{CV-10-2}
\begin{equation}
\Delta(\ell) \sim \xi_\ell^{-z}\sim \ell^{-\theta z},
\label{delsca}
\end{equation}
where $z=2$ is the dynamic exponent associated with the transition of
the homogeneous system, for both $d=1$ and $d=2$.  Within the TSS
framework, the scaling law of the singular part of the free-energy
density around the center of the trap is expected to behave
as~\cite{CV-10-2}
\begin{eqnarray}
&&F({\bf x},\mu,T,\ell) \approx \xi_\ell^{-(d+z)} {\cal
    F}({\bf x}_\ell,\,\mu_\ell,\,\tau_\ell),
\label{freee}\\
&&{\bf x}_\ell = {\bf x}/\xi_\ell,\quad
\mu_\ell = (\mu - \mu_{\rm vs})\,\xi_\ell^{y_\mu},
\quad \tau_\ell = T/ \Delta(\ell) .\qquad
\label{tssscavar}
\end{eqnarray}

Some issues concerning the equilibrium and out-of-equilibrium quantum
coherence and entanglement properties of trapped gases have been
already addressed~\cite{CV-10-2, CV-10-3, CV-10-4, CLM-15}.  In
particular, the study of the out-of-equilibrium quantum dynamics
requires the introduction of a further scaling variable related to the
time, i.e.
\begin{equation}
\theta_{t,\ell} = \Delta(\ell) \, t. 
\label{thetal}
\end{equation}
For example, when starting from the ground state of the initial trap,
and suddenly changing its size from $\ell_0$ to $\ell$, a generic
observable is expected to develop a dynamic TSS
behavior~\cite{CV-10-4}, such as
\begin{eqnarray}
&& O({\bf x};\mu,\ell_0,\ell,t) \approx \xi_{\ell_0}^{-y_o} {\cal
    O}({\bf x}_{\ell_0},\, \mu_{\ell_0}, \,\theta_{t,\ell_0}, \,
  \delta_\ell), \qquad
\label{scalbehNqu}\\
&&\xi_{\ell_0} \sim \ell_0^\theta,\quad \theta_{t,\ell_0}=\Delta(\ell_0) \, t,
\quad \delta_\ell = {\ell-\ell_0\over \ell_0},
\label{fu0sca}
\end{eqnarray}
where $y_o$ is the RG exponent controlling the scaling behavior of the
observable $O$ at the vacuum-to-superfluid transition.  The TSS limit
is obtained by taking the arguments of the scaling function ${\cal O}$
fixed.  This dynamic TSS Ansatz has been confirmed by the time
evolution of one-dimensional gases of impenetrable
bosons~\cite{CV-10-4}.

In the following we discuss the scaling properties of the work
fluctuations associated with quenches of the trap frequencies.  The
spectrum within a harmonic trap is discrete, thus the work fluctuation
probability can be straightforwardly defined as in Eq.~(\ref{pwdef}).
The dynamic TSS is the optimal framework to infer the trap-size
dependence of the work fluctuations after quenching the trap size. For
this purpose, analogously to the case of the DFSS of homogeneous
systems, we need to introduce a further scaling variable associated
with the work variable $W$.  We define
\begin{equation}
w_\ell = W/\Delta(\ell) \sim W\,\ell^{\,\theta z} .
\label{omegaldef}
\end{equation}
Using scaling arguments analogous to those exploited to investigate
finite-size effects in homogeneous systems, we arrive at the TSS Ansatz
\begin{equation}
P(W,\mu,\ell_0,\ell) \approx \Delta({\ell_0})\, {\cal P}
(w_{\ell_0},\,\mu_{\ell_0},  \,\delta_\ell).
\label{genpwscadtrep}
\end{equation}
As a consequence, the dynamic TSS behavior of the average work reads:
\begin{equation}
\langle W \rangle \approx \Delta(\ell_0) \,{\cal
  W}_1(\mu_{\ell_0},\,\delta_\ell).
\label{avwotss}
\end{equation}
The scaling behavior of the higher moments of the work fluctuations
can be easily obtained from Eq.~(\ref{genpwscadtrep}).

The dynamic TSS behavior of the work fluctuations can be checked by an
alternative derivation of the scaling behavior~(\ref{avwotss}) of the
work average.  Proceeding analogously to Sec.~\ref{fsscqt}, we can
write
\begin{equation}
\langle W \rangle = \langle 0_{\ell_0} | \,\sum_{\bf x} \big[ V({\bf
    x},\ell)-V({\bf x},\ell_0) \big] \,n_{\bf x} \,|
0_{\ell_0}\rangle.
\label{wdifftss}
\end{equation}
Then, using the TSS of particle density at the vacuum-to-superfluid
transition~\cite{CV-10-2},
\begin{equation}
\langle 0_{\ell_0} | n_{\bf x} | 0_{\ell_0}\rangle \approx
\xi_{\ell_0}^{-(d+z-y_\mu)} {\cal D}({\bf x}_\ell,\mu_\ell),
\label{desca}
\end{equation}
where ${\cal D}$ is a TSS function, we easily recover the dynamic TSS
reported in Eq.~(\ref{avwotss}), with
\begin{equation}
{\cal W}_1(\mu_{\ell_0},\delta_\ell) = - p \,\delta_\ell \int d^d {\bf
  x}\: |{\bf x}|^p \,{\cal D}({\bf x},\mu_{\ell_0}) .
\label{w1tss}
\end{equation}

Analogous studies can be performed at the $n=1$ Mott transition.
However, the corresponding dynamic behaviors is expected to be more
complicated.  We recall that the behavior around the $n=1$
Mott-to-superfluid transition of the homogeneous BH model without trap
is essentially analogous to that at the vacuum-to-superfluid
transition, because of the invariance under the particle-hole
exchange.  However, the particle-hole symmetry does not hold in the
presence of the trapping potential, and the asymptotic TSS dependence
becomes more involved~\cite{CV-10-2}.  This is essentially related to
the presence of level crossings at finite values of the trap size,
where the gap vanishes. The resulting trap-size dependence can be cast
in the form of a modulated TSS, that is a TSS controlled by the same
exponents as those at the low-density vacuum-to-superfluid transition,
but modulated by periodic functions of the trap size~\cite{CV-10-2}.
Analogous complications are expected to emerge also in dynamic
behaviors arising from quench protocols.  This issue however lies
outside the purpose of the present work and is left for future
investigations.

\section{Summary and outlook}
\label{conclu}

We have studied the scaling properties of the statistics of the work
done on a many-body system after a quench in proximity of a quantum
transition of any type. This has been done within a DFSS framework.
Close to a quantum transition, an asymptotic DFSS behavior emerges
from the interplay of the parameters involved in the quench protocol
and the size of the system.  In particular, we have considered a
generic Hamiltonian $H(\lambda) = H_c + \lambda H_p$, with
$[H_c,H_p]\neq 0$, and focused on a sudden change of the parameter
$\lambda$, assuming that the pre- and post-quench Hamiltonians remain
in the critical regime of a quantum transition. The DFSS limit is
defined as the large-size limit keeping appropriate scaling variables
fixed, associated with the Hamiltonian parameters, the temperature,
etc.  At CQTs these are ruled by suitable critical exponents and by
the RG dimension of the tuning parameter, with typical power-law
scaling behaviors.  On the other hand, at FOQTs they are dictated by
the size dependence of the energy gap, which is typically exponential,
but can also be power-law, depending on the type of boundary
conditions~\cite{CNPV-14}.  In our theoretical developments, we have
also kept into account the effect of finite temperatures (for systems
initially prepared into an equilibrium Gibbs ensemble), and discussed
how it is possible to consider generalized time-dependent protocols.

We stress the generality of the scaling arguments that we used to
develop the DFSS theory of the work fluctuations arising from quench
at quantum transitions.  As a consequence, they are expected to apply
to generic CQTs and FOQTs in any spatial dimension. The DFSS framework
allows us to infer the universal features of the scaling behaviors of
the work fluctuations, such as their power laws and corresponding
scaling functions (apart from trivial normalizations of the DFSS
variables).  Such predictions are expected to be independent of the
microscopic details of the models at hand, but only determined by a
few global properties shared by the universality class of the
transition at CQTs, and by the general features of the FOQT, such as
the fact that it may arise from a quasi-avoided two-level crossing in
the large-$L$ limit.

The predictions of the DFSS theory have been verified in two
paradigmatic quantum many-body systems, driven out of equilibrium by a
time variation of one of their characterizing parameters: the quantum
Ising and the Bose-Hubbard model.  For the Ising model, we considered
quenches associated with changes of a longitudinal magnetic
field. Depending on the value of the transverse field, the quench can
drive the system through either CQTs or FOQTs. We explicitly tested
our Ansatz by means of a numerical diagonalization of the Hamiltonian
at CQTs in one dimension, and also through a two-level truncation of
the spectrum at FOQTs.  For the Bose-Hubbard model, we numerically
studied quenches in the hard-core limit, driven by an external
constant field globally coupled to the bosonic modes, through the
vacuum-to-superfluid CQT.  We also generalized the DFSS framework for
the work fluctuations to describe particle systems confined by
inhomogeneous external potentials; in particular, we considered quench
protocols related to the variation of the size of the harmonic trap
confining the particle gas.

Issues related to work fluctuations are particularly relevant for
many-body systems of relatively small sizes, where they may be
meaningful and also experimentally detectable, while they are
conjectured to be largely suppressed in the thermodynamic
limit. Indeed, in that case, sizable fluctuations of the intensive
work density around its average value are expected to be extremely
rare~\cite{GS-12,GPGS-18}, thus hardly observable.  In this respect,
approaches based on DFSS frameworks are particularly suitable to infer
observable phenomena associated with quantum transitions in finite
(even quite small) many-body systems, and capture their universal
behaviors that are shared with a large class of models.  Remarkably,
our numerical results show that the DFSS behavior can be observed for
relatively small sizes: in some cases a limited number of spins
already displays the predicted asymptotic behavior.  Therefore, even
systems of modest size ($L$ order of $10$) may disclose the DFSS laws
of the work fluctuations derived in this paper.

Given the actual experimental interest for the nonequilibrium aspects
of the quantum dynamics of many-body systems, our results may be
particularly relevant for experimental investigations of the
properties of quantum work after quenches at quantum transitions.
Present-day quantum-simulation platforms have already demonstrated
their capability to reproduce and control the dynamics of quantum
Ising-like chains with $\sim 10$ spins. Ultracold atoms in optical
lattices~\cite{Simon-etal-11}, trapped
ions~\cite{Edwards-etal-10,Islam-etal-11,LMD-11,Kim-etal-11,Debnath-etal-16},
and Rydberg atoms~\cite{Labuhn-etal-16} may be promising candidates
where the emerging universality properties of the quantum many-body
physics can be tested with a minimal number of controllable objects.

\section*{References} %


\begin{thebibliography}{99}

\bibitem{BDZ-08} Bloch I, Dalibard J and Zwerger W,
  {\it Many-body physics with ultracold gases},
  2008 Rev. Mod. Phys. {\bf 80} 885

\bibitem{PSSV-11}
  Polkovnikov A, Sengupta K, Silva A and Vengalattore M,
  {\it Colloquium: Nonequilibrium dynamics of closed
  interacting quantum systems} 2011 Rev. Mod. Phys. {\bf 83} 863

\bibitem{HN-15} Nandkishore R and Huse D A, {\it Many body localization
  and thermalization in quantum statistical mechanics},
  2015 Annu. Rev. Condens. Matter Phys. {\bf 6} 15

\bibitem{Dziarmaga-10} Dziarmaga J, {\it Dynamics of a quantum phase
  transition and relaxation to a steady state},
  2010 Adv. Phys. {\bf 59} 1063
  
\bibitem{Jarzynski-11}
  Jarzynski C, {\it Equalities and inequalities: Irreversibility
  and the second law of thermodynamics at the nanoscale},
  2011 Annu. Rev. Condens. Matter Phys. {\bf 2} 329

\bibitem{Seifert-12} Seifert U, {\it Stochastic thermodynamics,
  fluctuation theorems and molecular machines},
  2012 Rep. Prog. Phys. {\bf 75} 126001

\bibitem{CHT-11} Campisi M, H\"anggi P and Talkner P,
  {\it Colloquium: Quantum fluctuation relations: Foundations and applications},
  2011 Rev. Mod. Phys. {\bf 83} 771

\bibitem{EHM-09} Esposito M, Harbola U and Mukamel S,
  {\it Nonequilibrium fluctuations, fluctuation theorems, and counting
  statistics in quantum systems}, 2009 Rev. Mod. Phys. {\bf 81} 1665

\bibitem{TLH-07}
  Talkner P, Lutz E and H\"anggi P, {\it Fluctuation theorems:
  Work is not an observable}, 2007 Phys. Rev. E {\bf 75} 050102(R)

\bibitem{TH-16}
  Talkner P and H\"anggi P,
  {\it Aspects of quantum work}, 2016 Phys. Rev. E {\bf 93} 022131

\bibitem{GPGS-18}
  Goold J, Plastina F, Gambassi A and Silva A,
  {\it The role of quantum work statistics in many-body physics},
  2018 arXiv:1804.02805

\bibitem{Silva-08} Silva A, {\it Statistics of the Work Done on a Quantum
  Critical System by Quenching a Control Parameter},
  2008 Phys. Rev. Lett. {\bf 101} 120603

\bibitem{DPK-08}
  Dorosz S, Platini T and Karevski D,
  {\it Work fluctuations in quantum spin chains},
  2008 Phys. Rev. E {\bf 77} 051120
  
\bibitem{Dorner-etal-12}
  Dorner R, Goold J, Cormick C, Paternostro M and Vedral V,
  {\it Emergent Thermodynamics in a Quenched Quantum Many-Body System},
  2012 Phys. Rev. Lett. {\bf 109} 160601

\bibitem{Mascarenhas-etal-14} Mascarenhas E, Braganc H, Dorner R,
  Franca Santos M, Vedral V, Modi K and Goold J, {\it Work and quantum
  phase transitions: Quantum latency}, 2014 Phys. Rev. E {\bf 89} 062103

\bibitem{MS-14} Marino J and Silva A, {\it Non-Equilibrium Dynamics of a
  Noisy Quantum Ising Chain: statistics of the work and
  prethermalization after a sudden quench of the transverse field},
  2014 Phys. Rev. B {\bf 89} 024303

\bibitem{ZT-15} Zhong M and Tong P, {\it Work done and irreversible
  entropy production in a suddenly quenched quantum spin chain with
  asymmetrical excitation spectra}, 2015 Phys. Rev. E {\bf 91} 032137

\bibitem{SD-15} Sharma S and Dutta A, {\it One- and two-dimensional
  quantum models: Quenches and the scaling of irreversible entropy},
  2015 Phys. Rev. E {\bf 92} 022108

\bibitem{Bayat-etal-16} Bayat A, Apollaro T J G, Paganelli S,
  De Chiara G, Johannesson H, Bose S and Sodano P, {\it Nonequilibrium
  critical scaling in quantum thermodynamics}, 2016 Phys. Rev. B {\bf 93}
  201106(R)

\bibitem{DL-08} Deffner S and Lutz E, {\it Nonequilibrium work
  distribution of a quantum harmonic oscillator},
  2008 Phys. Rev. E {\bf 77} 021128

\bibitem{GS-12} Gambassi A and Silva A, {\it Large Deviations and
  Universality in Quantum Quenches}, 2012 Phys. Rev. Lett. {\bf 109} 250602

\bibitem{SRH-14}
  Shchadilova Y E, Ribeiro P and Haque M,
  {\it Quantum Quenches and Work Distributions in Ultralow-Density Systems},
  2014 Phys. Rev. Lett. {\bf 112} 070601

\bibitem{SGLP-14}
  Sindona A, Goold J, Lo Gullo N and Plastina F,
  {\it Statistics of the work distribution for a quenched Fermi gas},
  2014 New J. Phys. {\bf 16} 045013

\bibitem{SGS-13}
  Sotiriadis S, Gambassi A and Silva A,
  {\it Statistics of the work done by splitting a one-dimensional quasicondensate},
  2013 Phys. Rev. E {\bf 87} 052129

\bibitem{SS-13}
  Smacchia P and Silva A, {\it Work distribution and edge
  singularities for generic time-dependent protocols in extended
  systems}, 2013 Phys. Rev. E {\bf 88} 042109

\bibitem{PS-14} P\'almai T and Sotiriadis S, {\it Quench echo and work
  statistics in integrable quantum field theories},
  2014 Phys. Rev. E {\bf 90} 052102

\bibitem{Palmai-15} P\'almai T, {\it Edge exponents in work statistics out
  of equilibrium and dynamical phase transitions from scattering
  theory in one-dimensional gapped systems}, 2015 Phys. Rev. B {\bf 92} 235433

\bibitem{BAKP-11}
  Bunin G, D'Alessio L, Kafri Y and Polkovnikov A,
  {\it Universal energy fluctuations in thermally isolated driven systems},
  2011 Nat. Phys. {\bf 7} 913
  
\bibitem{HPK-13} Heyl M, Polkovnikov A and Kehrein S, {\it Dynamical
  Quantum Phase Transitions in the Transverse-Field Ising Model},
  2013 Phys. Rev. Lett. {\bf 110} 135704

\bibitem{HSDL-08} Huber G, Schmidt-Kaler F, Deffner S and Lutz E,
  {\it Employing trapped cold ions to verify the quantum Jarzynski
  equality}, 2008 Phys. Rev. Lett. {\bf 101} 070403

\bibitem{Dorner-etal-13} Dorner R, Clark S R, Heaney L, Fazio R,
  Goold J and Vedral V, {\it Extracting Quantum Work Statistics and
  Fluctuation Theorems by Single-Qubit Interferometry},
  2013 Phys. Rev. Lett. {\bf 110} 230601

\bibitem{MDP-13} Mazzola L, De Chiara G and Paternostro M,
  {\it Measuring the characteristic function of the work distribution},
  2013 Phys. Rev. Lett. {\bf 110} 230602

\bibitem{CPV-14}
  Campostrini M, Pelissetto A and Vicari E,
  {\it Finite-size scaling at quantum transitions},
  2014 Phys. Rev. B {\bf 89} 094516
  
\bibitem{CNPV-14} Campostrini M, Nespolo J, Pelissetto A and Vicari E,
  {\it Finite-size scaling at first-order quantum transitions},
  2014 Phys. Rev. Lett. {\bf 113} 070402;
  {\it Finite-size scaling at first-order quantum transitions of
  quantum Potts chains}, 2015 Phys. Rev. E {\bf 91} 052103

\bibitem{PRV-18b} Pelissetto A, Rossini D and Vicari E, {\it Dynamic
  finite-size scaling after a quench at quantum transitions},
  2018 Phys. Rev. E {\bf 97} 052148

\bibitem{PRV-18}
  Pelissetto A, Rossini D and Vicari E,
  {\it Off-equilibrium dynamics driven by localized time-dependent
  perturbations at quantum phase transitions},
  2018 Phys. Rev. B {\bf 97} 094414

\bibitem{AFOV-08} Amico L, Fazio R, Osterloh A and Vedral V,
  {\it Entanglement in many-body systems},
  2008 Rev. Mod. Phys. {\bf 80} 517

\bibitem{CCD-09}
  {\em Entanglement entropy in extended systems},
  edited by Calabrese P, Cardy J and Doyon B,
  2009 J. Phys. A {\bf 42} 500301

\bibitem{TRHA-11}
  Tomasello B, Rossini D, Hamma A and Amico L,
  {\it Ground-state factorization and correlations with broken symmetry},
  2011 Eur. Phys. Lett. {\bf 96} 27002

\bibitem{DLLS-12}
  De Chiara G, Lepori L, Lewenstein M and Sanpera A,
  {\it Entanglement spectrum, critical exponents, and order parameters
  in quantum spin chains},
  2012 Phys. Rev. Lett. {\bf 109} 237208

\bibitem{CMC-13}
  Campbell S, Mazzola L, De Chiara G, Apollaro T J G, Plastina F,
  Busch Th. and Paternostro M,
  {\it Global quantum correlations in finite-size spin chains},
  2013 New J. Phys. {\bf 15} 043033

\bibitem{GH-13}
  Giampaolo S M and Hiesmayr B C,
  {\it Genuine multipartite entanglement in the XY model},
  2013 Phys. Rev. A {\bf 88} 052305

\bibitem{Bayat_2017}
  Bayat A,
  {\it Scaling of tripartite entanglement at impurity quantum phase transitions},
  2017 Phys. Rev. Lett. {\bf 118} 036102

\bibitem{DS-18}
  De Chiara G and Sanpera A,
  {\it Genuine quantum correlations in quantum many-body systems: a review of recent progress},
  2018 Rep. Prog. Phys. {\bf 81} 074002 

\bibitem{Gu-10}
  Gu S-J, {\it Fidelity approach to quantum phase transitions},
  2010 Int. J. Mod. Phys. B {\bf 24} 437

\bibitem{RV-18} Rossini D and Vicari E, {\it Ground-state fidelity at
  first-order quantum transitions}, 2018 arXiv:1807.01674

\bibitem{Zurek-03} Zurek W H, {\it Decoherence, einselection, and the
  quantum origins of the classical},
  2003 Rev. Mod. Phys. {\bf 75} 715

\bibitem{JP-09} Jacquod Ph and Petitjean C, {\it Decoherence,
  entanglement and irreversibility in quantum dynamical systems with
  few degrees of freedom}, 2009 Adv. Phys. {\bf 58} 67

\bibitem{V-18}
  Vicari E, {\it Decoherence dynamics of qubits coupled to
  systems at quantum transitions}, 2018 Phys. Rev. A {\bf 98} 052127

\bibitem{Barber-83} Barber M N, {\it Finite-size scaling, in Phase
  transitions and critical phenomena}, vol. 8, page 145, Domb C and
  Lebowitz J L eds. (1983 Academic Press, London)

\bibitem{Privman-90} {\em Finite Size Scaling and Numerical
  Simulations of Statistical Systems}, ed. Privman V (1990 World
  Scientific)

\bibitem{Sachdev-book} Sachdev S, {\em Quantum Phase Transitions},
  (1999 Cambridge University, Cambridge, England)

\bibitem{CPV-15} Campostrini M, Pelissetto A and Vicari E, {\it Quantum
  transitions driven by one-bond defects in quantum Ising rings},
  2015 Phys. Rev. E {\bf 91} 042123; {\it Quantum Ising chains with
  boundary terms}, 2015 J. Stat. Mech. P11015.

\bibitem{PRV-18c}
  Pelissetto A, Rossini D and Vicari E,
  {\it Finite-size scaling at first-order quantum transitions when boundary
  conditions favor one of the two phases},
  2018 Phys. Rev. E {\bf 98} 032124

\bibitem{ZJ-book} Zinn-Justin J, {\em Quantum Field Theory and
  Critical Phenomena}, fourth edition (2002 Clarendon Press, Oxford)

\bibitem{PV-02} Pelissetto A and Vicari E, 
  {\it Renormalization-group theory and critical phenomena},
  2002 Phys. Rep. {\bf 368} 549

\bibitem{Hasenbusch-10} Hasenbusch M, {\it A finite size scaling study of
  lattice models in the three-dimensional Ising universality class},
  2010 Phys. Rev. B {\bf 82} 174433

\bibitem{KPSV-16} Kos F, Poland D, Simmons-Duffin D and Vichi A,
  {\it Precision islands in the Ising and O($N$) models},
  2016 J. High Energy Phys. {\bf 08} 036

\bibitem{KP-17} Kompaniets M V and Panzer E, {\it Minimally subtracted
  six-loop renormalization of O($n$)-symmetric $\varphi^4$ theory and
  critical exponents}, 2017 Phys. Rev. D {\bf 96} 036016 

\bibitem{CHPV-00}
  Caselle M, Hasenbusch M, Pelissetto A and Vicari E,
  {\it Irrelevant operators in the two-dimensional Ising model},
  2002 J. Phys. A {\bf 35} 4861

\bibitem{Pfeuty-70}
  Pfeuty P, {\it The one-dimensional Ising model with a
  transverse field}, 1970 Ann. Phys. {\bf 57} 79

\bibitem{PF-83}
  Privman V and Fisher M E, {\it Finite-size effects at
  first-order transitions}, 1983 J. Stat. Phys. {\bf 33} 385

\bibitem{CJ-87} Cabrera G G and Jullien R, {\it Universality of
  Finite-Size Scaling: Role of the Boundary Conditions},
  1986 Phys. Rev. Lett. {\bf 57} 393; {\it Role of the boundary
  conditions in the finite-size Ising model}, 1987 Phys. Rev. B {\bf 35} 7062

\bibitem{Note}
  For systems with up to $L=12$ sites, we used an exact diagonalization approach.
  For larger sizes, in order to evaluate the ground state $|0_{\lambda_0}\rangle$
  of the pre-quench Hamiltonian, we employed a Lanczos-like algorithm
  (note that the first and the second moment of the work distribution can be found
  without evaluating the full spectrum of the post-quench Hamiltonian).
  See: Lehoucq R, Sorensen D and Yang C, {\em ARPACK Users Guide:
    Solution of Large Scale Eigenvalue Problems by Implicitly Restarted Arnoldi Methods},
  1997 Rice Univ. Press, Houston, USA

\bibitem{FWGF-89}
  Fisher M P A, Weichmann P B, Grinstein G and Fisher D S,
  {\it Boson localization and the superfluid-insulator transition},
  1989 Phys. Rev. B {\bf 40} 546

\bibitem{JBCGZ-98}
  Jaksch D, Bruder C, Cirac J I, Gardiner C W and Zoller P,
  {\it Cold bosonic atoms in optical lattices},
  1998 Phys. Rev. Lett. {\bf 81} 3108

\bibitem{CV-10-2} Campostrini M and Vicari E, {\it Quantum critical
  behavior and trap-size scaling of trapped bosons in a
  one-dimensional optical lattice},
  2010 Phys. Rev. A {\bf 81} 063614

\bibitem{RM-04} Rigol M and Muramatsu A, {\it Universal properties of
  hard-core bosons confined on one-dimensional lattices},
  2004 Phys. Rev. A {\bf 70} 031603(R);
  {\it Ground-state properties of hard-core bosons confined on one-dimensional optical lattices},
  2005 Phys. Rev. A {\bf 72} 013604

\bibitem{CV-10-4} Campostrini M and Vicari E, {\it Equilibrium and
  off-equilibrium trap-size scaling in 1D ultracold bosonic gases},
  2010 Phys. Rev. A {\bf 82} 063636

\bibitem{V-12} Vicari E, {\it Quantum dynamics and entanglement in
  one-dimensional Fermi gases released from a trap}, 2012 Phys. Rev. A {\bf
    85} 062324

\bibitem{CSC-13} Collura M, Sotiriadis S and Calabrese P,
  {\it Equilibration of a Tonks-Girardeau gas following a trap release},
  2013 Phys. Rev. Lett. {\bf 110} 245301; {\it Quench dynamics of a
  Tonks-Girardeau gas released from a harmonic trap},
  2013 J. Stat. Mech. P09025

\bibitem{CV-09} Campostrini M and Vicari E, {\it Critical behavior and
  scaling in trapped systems}, 2009 Phys. Rev. Lett. {\bf 102} 240601;
  2009 Phys. Rev. Lett. (E) {\bf 103} 269901; {\it Trap-size scaling in confined
  particle systems at quantum transitions}, 2010 Phys. Rev. A {\bf 81} 023606

\bibitem{CV-10-3}
  Campostrini M and Vicari E, {\it Bipartite quantum entanglement
  of one-dimensional lattice systems with a trapping potential},
  2010 J. Stat. Mech.: Theory Exp. P08020.

\bibitem{CLM-15}
  Calabrese P, Le Doussal P and  Majumdar S N,
  {\it Random matrices and entanglement entropy of trapped Fermi gases},
  2015 Phys. Rev. A {\bf 91} 012303

\bibitem{Simon-etal-11}
  Simon J, Bakr W S, Ma R, Tai M E, Preiss P M and Greiner M,
  {\it Quantum simulation of antiferromagnetic
  spin chains in an optical lattice}, 2011 Nature {\bf 472} 307

\bibitem{Edwards-etal-10}
  Edwards E E, Korenblit S, Kim K, Islam R, Chang M-S,
  Freericks J K, Lin G-D, Duan L-M and Monroe C, {\it Quantum
  simulation and phase diagram of the transverse-field Ising model with
  three atomic spins}, 2010 Phys. Rev. B {\bf 82} 060412(R)

\bibitem{Islam-etal-11}
  Islam R, Edwards E E, Kim K, Korenblit S, Noh C, Carmichael H,
  Lin G-D, Duan L-M, Joseph Wang C-C, Freericks J K and
  Monroe C, {\it Onset of a quantum phase transition with a trapped ion
  quantum simulator}, 2011 Nat. Commun. {\bf 2} 377

\bibitem{LMD-11}  Lin G-D, Monroe C and Duan L-M, {\it Sharp Phase
  Transitions in a Small Frustrated Network of Trapped Ion Spins},
  2011 Phys. Rev. Lett.  {\bf 106} 230402

\bibitem{Kim-etal-11}
  Kim K, Korenblit S, Islam R, Edwards E E, Chang M-S, Noh C,
  Carmichael H, Lin G-D, Duan L-M, Joseph Wang C-C,
  Freericks J K and Monroe C, {\it Quantum simulation of the transverse
  Ising model with trapped ions}, 2011 New J. Phys. {\bf 13} 105003

\bibitem{Debnath-etal-16}
  Debnath S, Linke N M, Figgatt C, Landsman K A, Wright K and Monroe C,
  {\it Demonstration of a small programmable quantum computer with atomic qubits},
  2016 Nature {\bf 536} 63

\bibitem{Labuhn-etal-16}  Labuhn H, Barredo D, Ravets S, de Leseleuc S,
  Macri T, Lahaye T and Browaeys A, {\it Tunable two-dimensional 
  arrays of single Rydberg atoms for realizing quantum Ising models},
  2016 Nature {\bf 534} 667

\end{thebibliography}
\end{document}